\pdfoutput=1 
\documentclass[
  aps,%
  prx,
 twocolumn,%
 groupedaddress,%
 superscriptaddress,%
  showpacs,%
 letterpaper,%
 amsfonts,%
 footinbib,%
  10pt,%
 floatfix,%
]{revtex4-2}
 
\usepackage{dcolumn} 
\usepackage{tikz}
\usetikzlibrary{positioning,arrows.meta}

\usepackage{hyperref}
\hypersetup{
    colorlinks=true,
    citecolor=blue,
    linkcolor=blue , 
    filecolor=cyan,      
    urlcolor=magenta,
    pdfcreator = {\LaTeX\ and \flqq hyperref\frqq},
}

\RequirePackage{graphicx,amsmath,amssymb,bm}
\usepackage{bbm}
\usepackage{float}
\usepackage{framed} 
\usepackage{amsthm}
\usepackage{caption}
\usepackage{subcaption}
\usepackage[normalem]{ulem}
\usepackage{comment}
\usepackage{slashed} 
\usepackage{upgreek}
\usepackage{xcolor}
\usepackage{enumitem}
\usepackage{tikz-3dplot}

\graphicspath{ {./Figs/} }

\newcommand{\1}{\mathbbm{1}}

\newcommand{\bbz}{\mathbb{Z}}

\newcommand{\shiftamount}{0.15}

\def\be{\begin{equation}}
\def\ee{\end{equation}} 
\def\bsh{\begin{shaded}}
\def\esh{\end{shaded}} 
\def\bpm{\begin{pmatrix}}
\def\epm{\end{pmatrix}}

\begin{document} 

\title{$\mathbb{Z}_2$ lattice gauge theories: fermionic gauging, transmutation, and Kramers-Wannier dualities} 

\author{Lei Su} 
\affiliation{Department of Physics and Pritzker School of Molecular Engineering, University of Chicago, Chicago, Illinois 60637, USA 
}

\begin{abstract} 
We generalize the gauging of $\mathbb{Z}_2$ symmetries by inserting Majorana fermions, establishing parallel duality correspondences for bosonic and fermionic lattice systems. Using this fermionic gauging, we construct fermionic analogs of $\mathbb{Z}_2$ gauge theories dual to the transverse-field Ising model, interpretable as Majorana stabilizer codes. We demonstrate a unitary equivalence between the $\mathbb{Z}_2$ gauge theory obtained by gauging the fermion parity of a free fermionic system and the conventional $\mathbb{Z}_2$ gauge theory with potentially nonlocal terms on the square lattice with toroidal geometry. This equivalence is implemented by a linear-depth local unitary circuit, connecting the bosonic and fermionic toric codes through a direction-dependent anyonic transmutation. The gauge theory obtained by gauging fermion parity is further shown to be  equivalent to a folded Ising chain obtained via the Jordan--Wigner transformation. We clarify the distinction between the recently proposed Kramers--Wannier dualities and those obtained by gauging the $\mathbb{Z}_2$ symmetry along a space-covering path. Our results extend naturally to higher-dimensional $\mathbb{Z}_2$ lattice gauge theories, providing a unified framework for bosonic and fermionic dualities and offering new insights for quantum computation and simulation.
\end{abstract} 
\maketitle

\tableofcontents

\section{Introduction}
The $\bbz_2$ lattice gauge theories dual to the transverse-field Ising model (TFIM) in various dimensions have been studied for decades \cite{savit1980, kapustin2014coupling}. In modern terminology, these theories can be obtained by gauging the spin-flip symmetry of the TFIM under the flatness condition.   
A solvable limit of the $\bbz_2$ lattice gauge theories with a standard Gauss law is known as toric codes \cite{kitaev2006anyons, dennisTopological2002}, which serve as canonical examples of topological quantum error-correcting codes. Their ground states exhibit long-range entanglement, topological degeneracy, and nontrivial anyonic excitations \cite{wen2004quantum}. There exists another class of $\bbz_2$ lattice gauge theories with a Gauss law that breaks the apparent lattice rotational symmetry, arising from exact bosonization \cite{bravyi2002fermionic, chen2018exact, chen2019bosonization}.  In these theories, the basic local excitations obey fermionic statistics, in contrast to the bosonic excitations of the gauge theory dual to the quantum Ising model. These gauge theories can be understood as resulting from gauging the fermion parity of a fermionic system~\cite{bhardwaj2017state, su2025bosonization}. In this work, we refer to the $\bbz_2$ gauge theories dual to the TFIM and to free fermionic systems as the Ising-dual and Majorana-dual $\bbz_2$ gauge theories, respectively, with the corresponding toric codes termed the bosonic and fermionic toric codes. The Majorana-dual theories are also known in the literature as shadow theories~\cite{bhardwaj2017state}. We develop  a parallel duality framework for both bosonic and fermionic lattice systems within the Hamiltonian formalism by extending conventional bosonic gauging, i.e., coupling to bosonic flat gauge spins, to fermionic gauging implemented by inserting Majorana fermions, a procedure shown to be equivalent to diagonal gauging after stacking with a fermionic system.

The bosonic and fermionic toric codes share the same ground-state structure but host distinct anyonic excitations. It is sometimes assumed that the anyons in these two codes can be identified by a simple permutation. We show, however, that the correspondence is more subtle. A linear-depth local unitary maps the bosonic and fermionic toric codes to each other, but it relates anyons in a nontrivial manner that can be interpreted as a direction-dependent transmutation, which also accounts for the rotational symmetry breaking in the fermionic case.

A distinctive feature of the Majorana-dual $\bbz_2$ gauge theories is the existence of generalized Kramers--Wannier (KW) self-dualities on regular lattices \cite{su2025bosonization}. These KW self-dualities are derived from minimal translations of the Majorana lattices and are direction-dependent, differing from those obtained via gauging higher-form symmetries \cite{choi2022noninvertible, kaidi2022kramers} or subsystem symmetries \cite{caoSubsystem2023}. We will show that they are inherited from, but distinct from, the 1D counterpart. 

The paper is structured as follows. In Sec.~\ref{sec_gauging}, we present the duality web between bosonic and fermionic systems via bosonic and fermionic gauging. We first illustrate the procedure using the 1D duality between the TFIM and the Majorana chain in Sec.~\ref{sec_gauging1d}, before turning to more interesting 2D generalizations in Sec.~\ref{sec_gaugingTFIM} and \ref{sec_gaugingMaj}. Starting in Sec.~\ref{sec_mapping}, we focus on bosonic gauge theories, relating the Majorana-dual gauge theory on a square lattice to the Ising-dual gauge theory by mapping both to ancilla-decorated Ising models. The direction-dependent flux attachment that accounts for the anyon transmutation is discussed in Sec.~\ref{sec_transmutation}. In Sec.~\ref{sec_jw}, we show that the gauge theory obtained by gauging fermion parity is equivalent to a folded Ising chain obtained via the Jordan--Wigner (JW) transformation, and we discuss the uniqueness of the generalized KW self-dualities. We extend the analysis to 3D in Sec.~\ref{sec_3d} and conclude in Sec.~\ref{sec_conclusion}. Additional details are provided in the Appendixes.

\begin{figure}[tb]
\includegraphics[width=1\linewidth]{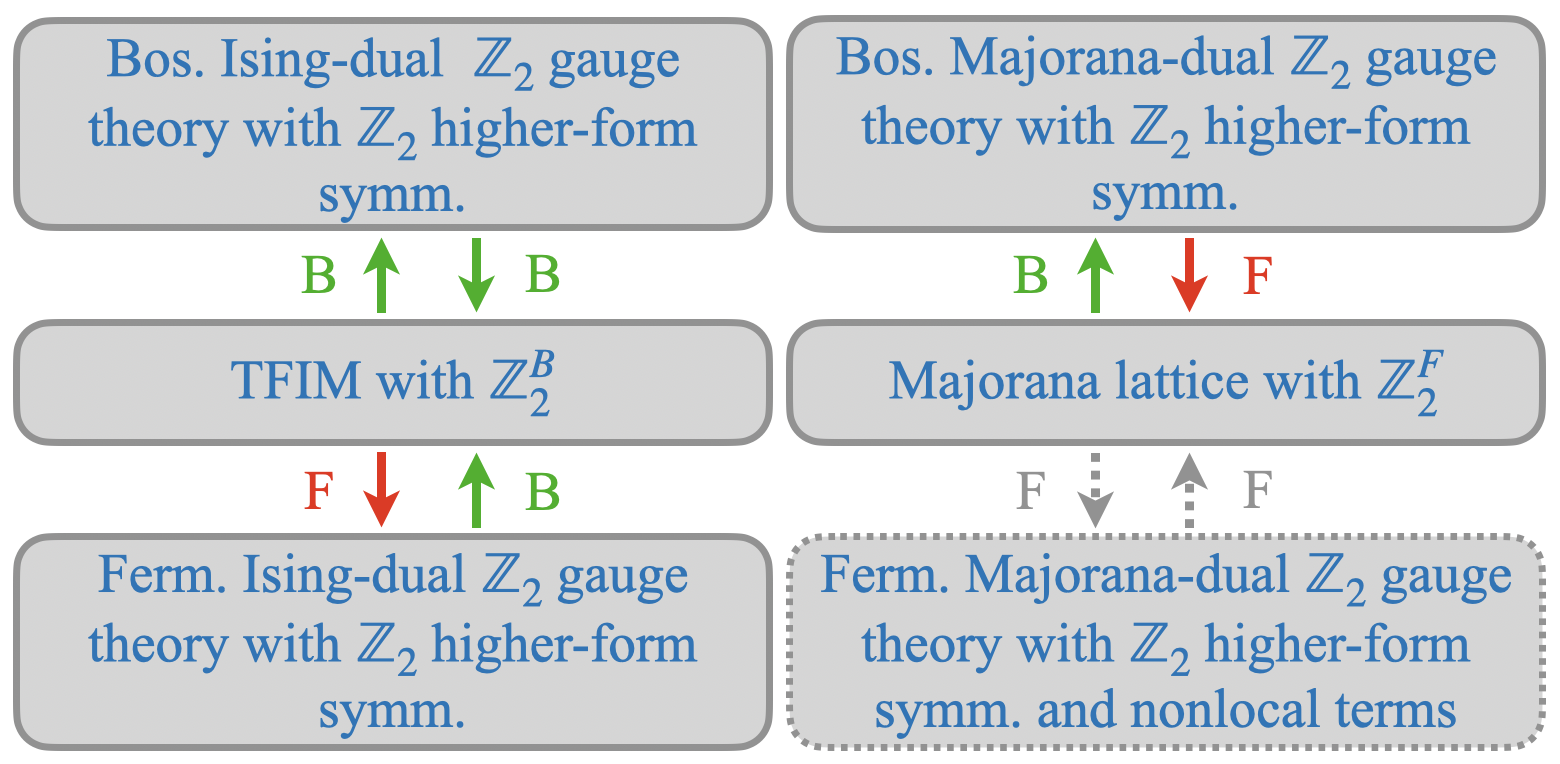}
\centering
\caption{Gauging and dualities of bosonic and fermionic systems. Green (red) arrows indicate bosonic (fermionic) gauging. Gray (dotted) arrows indicate fermionic gauging via a nonlocal disentangling unitary to and from a fermionic gauge theory with nonlocal terms.}
\label{fig:relation}
\end{figure}

\section{Gauging and dualities }
\label{sec_gauging}
Given a quantum Ising model with a global $\bbz_2^B$ spin-flip symmetry, the (bosonic) dual theory can be obtained by gauging this symmetry (see Fig.~\ref{fig:relation}), following the standard procedure of minimally coupling to flat gauge spins and applying a disentangling unitary transformation. The disentangling unitary, denoted $U_B$, is a tensor product of controlled gates. We refer to this procedure as bosonic gauging in what follows. The resulting dual theory in $n$ spatial dimensions is a $\bbz_2$ gauge theory with a dual $(n{-}1)$-form symmetry, whose charges are $(n{-}1)$-D \cite{gaiotto2015generalized}. Gauging this dual higher-form symmetry reconstructs the original Ising models, with the $\bbz_2^B$ spin-flip symmetry identified as the dual symmetry. The disentangling unitary remains identical to $U_B$, and this duality construction applies to general polyhedral decompositions of space. 

Analogous operations exist for parity-even fermionic systems. In $n$-D, gauging the $\bbz_2^F$ fermion parity yields a (bosonic) $\bbz_2$ gauge theory with a dual $(n{-}1)$-form symmetry, implemented through a disentangling unitary $U_F$~\cite{su2025bosonization}. The essential difference from the Ising case is that the ordering of the controlled gates in $U_F$ becomes crucial due to fermionic anti-commutation relations. The dual $(n{-}1)$-form symmetry is generated by loops of emergent fermions, which carry a nontrivial 't Hooft anomaly and  cannot be gauged without specifying spin structures \cite{gaiotto2016spin}. Nevertheless, this higher-form symmetry can be gauged to reconstruct the original system either by stacking with a physical fermionic system and condensing the composite \cite{gaiotto2016spin, bhardwaj2017state, aasenFermion2019}, or by inserting  Majorana fermions  \cite{chen2021disentangling}.  Using the (inverse) disentangling unitary $U_F$, we can show explicitly that these fermion-condensation approaches are equivalent. In fact, we demonstrate below that fermionic gauging is more general and does not rely on anomalies, and that it is a neat language to repackage the composite procedure of stacking with a fermionic insulator and gauging the diagonal $\bbz_2$ symmetry, or fermion condensation.    

\subsection{Gauging and dualities of TFIM and Majorana chain in 1D}
\label{sec_gauging1d}
\begin{figure}[b]
\includegraphics[width=0.55\linewidth]{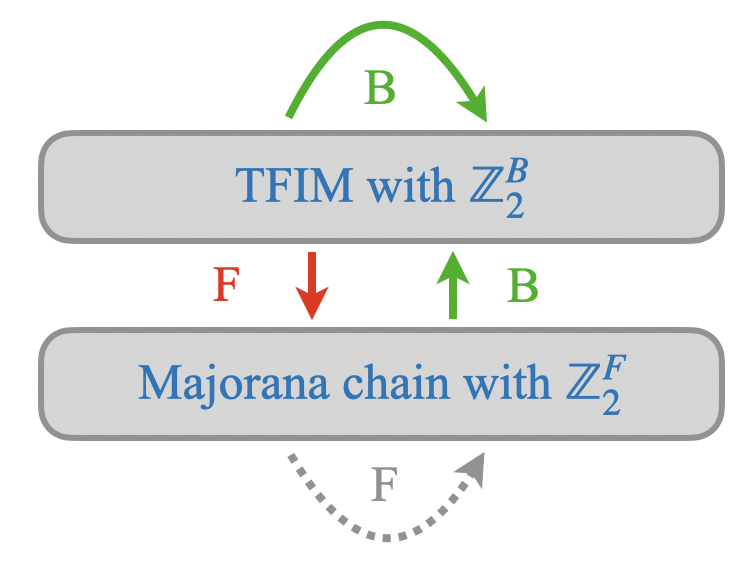}
\centering
\caption{Duality between the 1D TFIM and Majorana chain. Green (red) arrows indicate bosonic (fermionic) gauging. Gray (dotted) arrows indicate fermionic gauging via a nonlocal disentangling unitary.}
\label{fig:1d}
\end{figure}

To illustrate the idea, we revisit the classic duality between the 1D TFIM and the Majorana chain with periodic boundary conditions. In 1D, the six systems shown in Fig.~\ref{fig:relation} collapse into two and become connected, as shown in Fig.~\ref{fig:1d}.  Both gauging the $\bbz_2^B$ symmetry of the TFIM and gauging the fermion parity $\bbz_2^F$ of the Majorana chain by inserting gauge spins on the edges lead to the TFIM, corresponding respectively to the KW duality and the JW transformation. 

\subsubsection{Bosonic gauging of $\bbz_2^B$ in TFIM}
Given the Hamiltonian of the TFIM on a periodic chain, 
\be 
H_B = - J \sum_j Z_j Z_{j+1} - g\sum_j X_j,
\label{eq:Ham_Ising}
\ee 
we can follow the standard gauging scheme by inserting gauge spins on the edges to perform a KW duality transformation. The Gauss law is 
\be 
X_j \sigma_{j-1/2}^z \sigma_{j+1/2}^z =1,
\ee  
and the minimally coupled Hamiltonian is 
\be 
\tilde{H}_B = -J \sum_j  Z_j \sigma_{j+1/2}^x  Z_{j+1} - g\sum_j X_j.
\ee  
The unitary transformation under
\be 
U_B = \prod_j \left(P_{j+1/2}^+ +P_{j+1/2}^- Z_j Z_{j+1} \right),
\ee 
where $P_{j+1/2}^{\pm} = (1 \pm \sigma_{j+1/2}^z )/2$, 
maps the Gauss law to $X_j =1$ and the Hamiltonian to 
\be 
\tilde{H}'_B = - J \sum_j  \sigma^x_{j+1/2} - g\sum_j \sigma_{j-1/2}^z X_j  \sigma_{j+1/2}^z.  
\ee    
Plugging in $X_j =1$, we obtain the dual TFIM
\be 
H'_B = - J \sum_j  \sigma^x_{j+1/2} - g\sum_j \sigma_{j-1/2}^z  \sigma_{j+1/2}^z.  
\ee   

\subsubsection{Bosonic gauging of $\bbz_2^F$ in Majorana chain}
We can gauge the fermion parity $\bbz_2^F$ of the Majorana chain in a similar way. The Hamiltonian is 
\be 
H_F = - J \sum_j  i \gamma_{j} \gamma'_{j+1}  - g \sum_j i \gamma'_{j} \gamma_{j} . 
\ee  
Insert gauge spins on the edges and impose the Gauss law
\be 
(-1)^{F_j} \sigma_{j-1/2}^z \sigma_{j+1/2}^z =1,
\ee  
where $(-1)^{F_j} \equiv i \gamma'_j \gamma_j$. The minimally coupled Hamiltonian is 
\be 
\tilde{H}_F = -J \sum_j  \sigma_{j+1/2}^x  i \gamma_{j} \gamma'_{j+1} - g\sum_j (-1)^{F_j}.
\ee  
Using the disentangling unitary
\be 
U_F= \prod_j \left[P_{j+1/2}^+ +P_{j+1/2}^- (i \gamma_j \gamma'_{j+1}) \right],
\ee 
where $P_{j+1/2}^{\pm} = (1 \pm \sigma_{j+1/2}^z )/2$, to map the Gauss law to $(-1)^{F_j} =1$,  we obtain the TFIM
\be 
H'_B = - J \sum_j  \sigma^x_{j+1/2} - g\sum_j \sigma_{j-1/2}^z  \sigma_{j+1/2}^z.  
\ee   
This transformation corresponds to the traditional JW transformation (see Appendix~\ref{app_jw}).

\subsubsection{Fermionic gauging of $\bbz_2^B$ TFIM}
Consider again the Hamiltonian $H_B$ of the TFIM in Eq.~(\ref{eq:Ham_Ising}). The $\bbz_2^B$ symmetry generated by $\prod_j X_j$ in the TFIM can also be viewed as an anomalous symmetry. Indeed, the string operator $\prod_{j<k} Y_j X_{j+1}\cdots X_{k-1} Y_k$ is the bosonized representation of a fermionic bilinear. Thus, $\prod_j X_j$ can be regarded as the fermion loop obtained by annihilating the two ends of the string operator around the periodic chain. This becomes more transparent if we rewrite $\prod_j X_j \sim \prod_j (Z_{j-1} Z_jX_j)$.  The corresponding operator, $\prod_j (i\gamma'_j\gamma_j)(i\gamma_j \gamma'_{j+1})$, is a concatenated product of fermion bilinears forming a closed loop whose value depends on the boundary condition, i.e., the spin structure. This symmetry can be gauged by inserting Majorana fermion pairs on the edges and imposing the Gauss law  
\be 
Z_{j-1}Z_j (i\gamma_{j-1/2} \gamma'_{j+1/2})X_j (i\gamma'_{j+1/2} \gamma_{j+1/2}) =1.
\ee 
The minimally coupled Hamiltonian is 
\be 
\tilde{H}_B = -J \sum_j (-1)^{F_{j+1/2}} Z_j Z_{j+1} - g\sum_j (-1)^{F_{j+1/2}} X_j,
\ee 
where $(-1)^{F_{j + 1/2}} \equiv i\gamma'_{j+ 1/2} \gamma_{j+ 1/2}$. The unitary transformation under 
\be 
U_0 = \prod_j \left({\cal{P}}_{j+1/2}^+ +{\cal{P}}_{j+1/2}^- Z_j \right),
\ee 
where ${\cal{P}}_{j+1/2}^{\pm} = [1 \pm (-1)^{F_{j+1/2}} ]/2$, 
maps the Gauss law to $X_j (i\gamma_{j-1/2} \gamma'_{j+1/2})=1$ and the Hamiltonian to 
\be 
\tilde{H}'_B = - J \sum_j (-1)^{F_{j+1/2}} Z_j  Z_{j+1} - g\sum_j X_j.  
\ee    
The disentangling unitary 
\be 
U_F = \prod_j \left[  P^{+}_j + P^-_j (i \gamma_{j-1/2} \gamma'_{j+1/2}) \right],
\ee 
where $P_j^{\pm} = (1 \pm Z_j )/2$, maps 
\be
\begin{split}
X_j & \to X_j (i \gamma_{j-1/2} \gamma'_{j+1/2}), \\ 
(-1)^{F_{j+1/2}} & \to (-1)^{F_{j+1/2}} Z_j Z_{j+1}. 
\end{split} 
\ee
Plugging $X_j =1$, we obtain
\be 
H_F = - J \sum_j i \gamma'_{j+1/2} \gamma_{j+1/2}   - g \sum_j i \gamma_{j-1/2} \gamma'_{j+1/2}. 
\ee 
Fermionic gauging starting with $X_j (i\gamma_{j-1/2} \gamma'_{j+1/2})=1$ is also discussed in Ref.~\cite{aksoyLiebSchultzMattis2024}.

\subsubsection{Gauging $\bbz_2^B$ in TFIM with stacking}
This fermionic gauging process can be shown to be equivalent to diagonal gauging after stacking the system with
a free fermionic system. We stack a fermionic system 
\be 
H'_F = -\kappa\sum i\gamma'_{j+1/2}\gamma_{j+1/2}
\ee 
to the TFIM, where $\kappa$ is taken to be large. We then gauge the diagonal $\bbz_2$ of $\bbz_2^B \times \bbz_2^F$ by imposing the Gauss law
\be X_j (i\gamma_{j-1/2} \gamma'_{j+1/2}) = \sigma^z_{j-1/2} \sigma^z_{j+1/2}\ee
on the Hamiltonian
\be
\begin{split}
 \tilde{H}_F  =& - J \sum_j Z_j \sigma_{j+1/2}^x Z_{j+1} - g \sum_j X_j   \\
 &  -\kappa \sum_j (-1)^{F_{j+1/2}} \sigma^x_{j+1/2}.   
\end{split} 
\ee
Compared to the first approach, the new Gauss law is relaxed by additional $\sigma^z$.
Note that if we shift the Majorana fermions in $H'_F$ by a minimal Majorana translation and stack a Majorana chain, also referred to as an invertible fermionic topological order \cite{ji2020top}, then the result is also shifted by the minimal translation. We now perform a unitary transformation using
\be 
U'_F = \prod_j \left[  P^{+}_j + P^-_j (i \gamma_{j-1/2} \gamma'_{j+1/2})( \sigma^z_{j-1/2} \sigma^z_{j+1/2})\right],
\ee 
where $P_j^{\pm} = (1 \pm Z_j )/2$. The Hamiltonian becomes
\be 
\begin{split}
  &\tilde{H}'_F = \\
  & - J \sum_j  \sigma_{j+1/2}^x 
  - g\sum_j \sigma^z_{j-1/2}  (i \gamma_{j-1/2} \gamma'_{j+1/2}) \sigma^z_{j+1/2}\\
  &-\kappa \sum_j (-1)^{F_{j+1/2}} \sigma^x_{j+1/2}  .  
\end{split}
\ee 
Another disentangling unitary transformation under 
\be 
U''_F = \prod_j \left[  \tilde{P}^{+}_{j+1/2} + \tilde{P}^-_{j+1/2} (-1)^{F_{j+1/2}} \right],
\ee 
where $\tilde{P}^{\pm}_{j+1/2} = (1\pm \sigma_{j+1/2}^z)/2$, maps the Hamiltonian to 
\be 
H_F = - J \sum_j  i \gamma'_{j+1/2}  \gamma_{j+1/2}   - g \sum_j  i \gamma_{j-1/2} \gamma'_{j+1/2},  
\ee  
where we have taken $\kappa \to \infty$ and used $\sigma^x_{j+1/2} =1$. Thus, we have recovered the same Majorana chain obtained above. In other words, two gauging schemes are equivalent.

\subsubsection{Fermionic gauging of $\bbz_2^F$ in Majorana chain} 
\label{sec_fermionic_analog}
We can also gauge the fermion parity $\bbz_2^F$ of the Majorana chain by inserting Majorana fermions and recover itself at the cost of using a nonlocal disentangling unitary circuit (indicated by the gray arrows in Fig.~\ref{fig:relation}). It can be regarded as the fermionic analog of the KW duality in the 1D TFIM, previously interpreted as stacking a Majorana chain \cite{ji2020top}. 
Starting with the Majorana chain, we insert Majorana fermions on the edges and impose the Gauss law as follows:
\be 
(-1)^{F_j} i\tilde{\gamma}_{j-1/2} \tilde{\gamma}'_{j+1/2} =1.
\ee 
The minimally coupled Hamiltonian becomes
\be 
\tilde{H}_F = - J\sum_{j} (-1)^{\tilde{F}_{j+1/2}} i\gamma_j \gamma'_{j+1}  - g \sum_j  i \gamma'_j \gamma_j.
\ee
To disentangle the original Majorana fermions, we can use the following unitary 
\be 
U'_F = \prod_j \left( \tilde{{\cal{P}}}_{j}^+ + \tilde{{\cal{P}}}_{j}^- \gamma'_j \right),
\ee 
where $\tilde{{\cal{P}}}_j^{\pm} = (1 \pm i\tilde{\gamma}_{j-1/2} \tilde{\gamma}'_{j+1/2} )/2$ and terms with smaller $j$ are placed to the right. We may also attach  fermion parity strings to $\gamma'_j$ to enforce the commutativity of the factors. Under the nonlocal unitary transformation of $U'_F$, 
\be
\begin{split}
  (-1)^{F_j} & \to  (-1)^{F_j} i\tilde{\gamma}_{j-1/2} \tilde{\gamma}'_{j+1/2}, \\ 
(-1)^{\tilde{F}_{j+1/2}} i\gamma_j \gamma'_{j+1} &\to (-1)^{\tilde{F}_{j+1/2}} (-1)^{F_j}.  
\end{split}
\ee 
Plugging the decoupled Gauss law $(-1)^{F_j} =1$, we obtain 
\be 
H'_F = - J\sum_j i \tilde{\gamma}'_{j+1/2}  \tilde{\gamma}_{j+1/2} - g \sum_j i\tilde{\gamma}_{j-1/2} \tilde{\gamma}'_{j+1/2}.
\ee 
At $J =g$, the Majorana chain is invariant, up to a minimal Majorana translation, under the fermionic gauging. Thus, the process leads to a fermionic analog of the KW self-duality in the 1D TFIM. Similar to the case of gauging $\bbz_2^B$ of the TFIM, it is straightforward to show that gauging with stacking is equivalent to the fermionic gauging.

\subsection{Gauging and dualities of TFIM and its duals in 2D}
\label{sec_gaugingTFIM}
\begin{figure}[tb]
    \centering
    \includegraphics[width=0.85\linewidth]{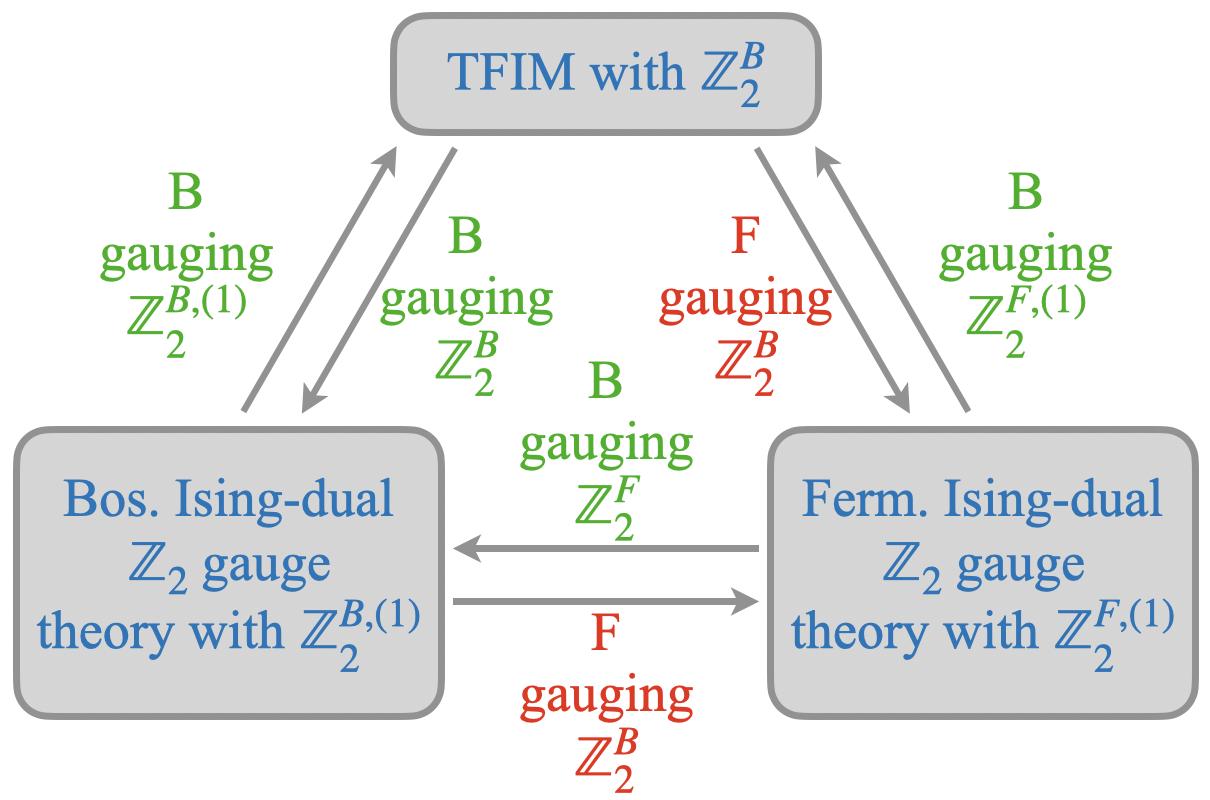}
    \caption{Duality triangle between the TFIM and its bosonic and fermionic dual gauge theories. $\bbz_2^B$ denotes the spin-flip symmetry, and $\bbz_2^F$ the fermion parity. $\bbz_2^{B,(1)}$ and $\bbz_2^{F,(1)}$ represent the corresponding 1-form symmetries. ``B gauging" and ``F gauging" refer to bosonic and fermionic gauging, respectively. }
    \label{fig:triangle}
\end{figure}

We now show that the discussion in 1D extends naturally to higher dimensions. For concreteness, we focus on the square lattice, though the procedure generalizes straightforwardly to general polyhedral decompositions of space in any dimension. We first examine the dualities between the TFIM and its dual gauge theories captured by the duality triangle shown in Fig.~\ref{fig:triangle}.

\begin{figure}[b]
    \centering 

\begin{tikzpicture}[scale=0.8, line cap=round, line join=round]

  \def\Nx{3}     
  \def\Ny{3}     
  \def\a{1.0}    

  \foreach \i in {0,...,\Ny}{
    \draw[line width=0.6pt] (0,\i*\a) -- (\Nx*\a,\i*\a);
  }
  \foreach \j in {0,...,\Nx}{
    \draw[line width=0.6pt] (\j*\a,0) -- (\j*\a,\Ny*\a);
  }

\filldraw[black] (1.5*\a,1.5*\a) circle (0.05);

\node[right] at (1.5*\a,1.3*\a){\large $r$};

\draw[red, very thick] (\a,\a) -- (\a,2*\a);
\draw[red, very thick] (\a,\a) -- (2*\a,\a);
\draw[red, very thick] (2*\a,\a) -- (2*\a,2*\a);
\draw[red, very thick] (\a,2*\a) -- (2*\a,2*\a);

\node[left] at (1*\a,2.2*\a){\large $v$};
\node[left] at (0.5*\a, 3.4*\a ) {\large $(a)$};

\pgfmathsetmacro{\xshift}{\Nx*\a+2*\a} 

  \foreach \i in {0,...,\Ny}{
    \draw[line width=0.6pt] (\xshift+0,\i*\a) -- (\xshift+\Nx*\a,\i*\a);
  }
  \foreach \j in {0,...,\Nx}{
    \draw[line width=0.6pt] (\xshift+\j*\a,0) -- (\xshift+\j*\a,\Ny*\a);
  }

\filldraw[black] (\xshift+1.3*\a,1.5*\a) circle (0.05);
\filldraw[black] (\xshift+1.7*\a,1.5*\a) circle (0.05);
\filldraw[black] (\xshift+2.3*\a,1.5*\a) circle (0.05);
\filldraw[black] (\xshift+2.7*\a,1.5*\a) circle (0.05);
\filldraw[black] (\xshift+1.3*\a,2.5*\a) circle (0.05);
\filldraw[black] (\xshift+1.7*\a,2.5*\a) circle (0.05);

\node[right] at (\xshift+1.1*\a,1.25*\a){\large $\gamma'$};
\node[right] at (\xshift+1.5*\a,1.2*\a){\large $\gamma$};

\draw[-Stealth, orange, thick] (\xshift+3.3*\a,2*\a)  -- (\xshift+3.3*\a,3*\a) ;

\draw[-Stealth, orange, thick] (\xshift+3.3*\a,2*\a)  -- (\xshift+4.3*\a,2*\a) ;

\node[above, orange] at (\xshift+2.5*\a,1.9*\a){\large $1$};
\node[left, orange] at (\xshift+3.1*\a,2.5*\a){\large $2$};

\draw[red, very thick] (\xshift+\a,\a) -- (\xshift+\a,2*\a);
\draw[red, very thick] (\xshift+\a,\a) -- (\xshift+2*\a,\a);
\draw[red, very thick] (\xshift+2*\a,\a) -- (\xshift+2*\a,2*\a);
\draw[red, very thick] (\xshift+\a,2*\a) -- (\xshift+2*\a,2*\a);

\draw[-Stealth, cyan, thick] (\xshift+1.3*\a,1.5*\a)  -- (\xshift+1.7*\a,2.5*\a) ;

\draw[-Stealth, cyan, thick] (\xshift+2.3*\a,1.5*\a)  -- (\xshift+1.7*\a,1.5*\a) ;

\node[left] at (\xshift+1*\a,2.2*\a){\large $v$};
\node[left] at (\xshift+0.5*\a, 3.5*\a ) {\large $(b)$};
\end{tikzpicture}
\caption{Gauging of the TFIM (a) and the Majorana lattice (b). The fermionic gauging of (a) is shown in Fig.~\ref{fig:2d}(c). In (b), two Majorana fermions, $\gamma'$ and $\gamma$, are placed within each plaquette. The cyan arrows denote two fermion bilinears which, together with the local parity operator, generate the parity-even fermionic algebra. To obtain the operators displayed in Fig.~\ref{fig:2d}(b), the ordering of edges in the unitary $U_F$ [Eq.~(\ref{eq:gaugingU2})] is specified by the orange numbers, increasing from left to right and bottom to top, as indicated by the orange arrows.}
\label{fig:bos_gauging}
\end{figure}

\subsubsection{Bosonic gauging of $\bbz_2^B$ in TFIM}
We first review the duality transformation by bosonic gauging. Consider the 2D TFIM on the square lattice with periodic boundary conditions. We place an Ising spin or a qubit at the center of each plaquette $r$ [Fig.~\ref{fig:bos_gauging}(a)]. The Hamiltonian is given by  
\be 
H_B = - J \sum_{r} \sigma^z_r \sigma^z_{r+\hat{x}} + \sigma^z_r \sigma^z_{r+\hat{y}} - g \sum_r  \sigma^x_r,
\label{eq:TFIM}
\ee 
where $\hat{x}, \hat{y}$ are the lattice vectors. The global $\bbz_2^B$ symmetry generated by $\prod_r \sigma^x_r$ can be gauged by placing the gauge spins on the edges, as indicated in red in Fig.~\ref{fig:bos_gauging}. We then impose the Gauss law for each plaquette
\be
 \sigma^x_r \prod_{e \subset \partial r} Z_e  =1.
\ee 
Using the minimal coupling procedure, we can write down the gauged Hamiltonian as
\be 
\tilde{H}_B = - J \sum_{r} \sigma^z_r X_{e_{r, \hat{x}}} \sigma^z_{r+\hat{x}} + \sigma^z_r X_{e_{r, \hat{y}}} \sigma^z_{r+\hat{y}} - g \sum_r  \sigma^x_r,
\ee 
where $X$ plays the role of a $\bbz_2$-valued discrete gauge field. We use $e_{r, \pm \hat{x}}$ $e_{r, \pm \hat{y}}$ to denote the edge of the plaquette $r$ in the $\pm \hat{x}$ or $\pm \hat{y}$ direction.  The flatness condition requires that the four spins incident on any vertex $v$ satisfy 
\be 
\prod_{\partial e \supset v} X_e =1. 
\label{eq:flatness}
\ee 
It forces the magnetic flux passing through $v$ to vanish, and there is no Aharonov--Bohm phase for $\sigma^z$ when taking different paths around a (dual) plaquette. Thus, the flatness condition removes the ambiguity associated with minimal coupling.

\begin{figure}[tb]
    \centering
    \includegraphics[width=0.65\linewidth]{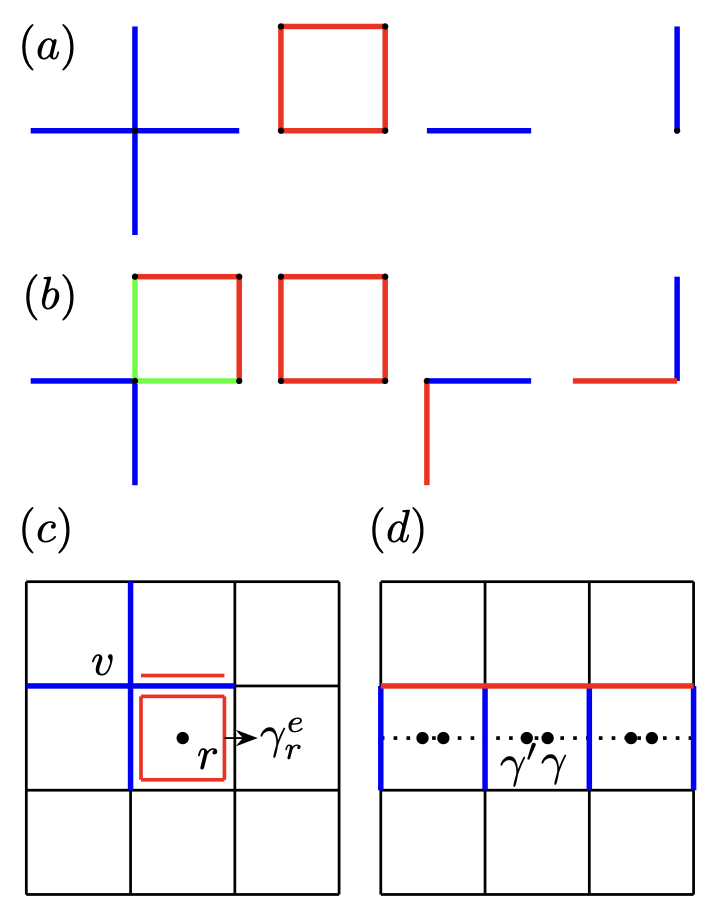}
    \caption{(a) Operators in the bosonic and fermionic Ising-dual gauge theories on the square lattice. In the bosonic (fermionic) theory, blue and red lines represent $X, Z$ [$(-1)^{F}, \gamma$], respectively. The first operator corresponds to the Gauss law. (b) Operators in the bosonic Majorana-dual theory [Eq.~(\ref{eq:MajoranaDual})]. Green lines represent $Y$. The overall minus sign in the Gauss-law operator is omitted. (c) Fermionic gauging of the TFIM. Two Majorana fermions (red lines) are assigned to an edge, and four Majorana fermions associated with a plaquette form a plaquette-like operator. (d) Fermonic gauging of the  $\bbz_2^{B, (1)}$ 1-form symmetry of the bosonic Majorana-dual theory.  The 1-form symmetry along the horizontal direction is generated by the product of $X$ and $Z$. A pair of Majorana fermions are introduced on each plaquette. }
    \label{fig:2d}
\end{figure}

Next, we perform a unitary transformation to disentangle the matter, i.e., the original spins, from the gauge spins. The local unitary circuit to implement the action is given by
\be 
U_B = \prod_e (P_e^+ + P_e^- \sigma^z_{r^e_1} \sigma^z_{r^e_2}).
\label{eq:gaugingU}
\ee 
Here, $P_e^{\pm} = (1\pm Z_e)/2$ are the projection operators for $Z$ on edge $e$. $r^e_{1,2}$ are the two plaquettes sharing the edge $e$. The ordering of $e$ does not matter because all factors commute. The adjoint action of this operator maps $\sigma_r^x$ to $\prod_{e \subset \partial r} Z_e$ and thus 
\be
\sigma^x_r \prod_{e \subset \partial r} Z_e \to \sigma^x_r = 1.
\ee 
It also maps $X_e$ to $\sigma^z_{r^e_1} X_e \sigma^z_{r^e_2}$, and thus
\be 
\sigma^z_{r^e_1} X_e \sigma^z_{r^e_2} \to  X_e,
\ee 
while the flatness condition remains invariant. Consequently, the final Hamiltonian, after setting $\sigma^x_r = 1$, becomes 
\be 
H'_B = - J \sum_e X_e - g \sum_r \prod_{e \subset \partial r} Z_e, 
\label{eq:z2gaugetheory}
\ee 
with the new Gauss law
\be 
\prod_{\partial e \supset v} X_e =1. 
\label{eq:gaugingGauss}
\ee 
This is precisely the bosonic Ising-dual $\bbz_2$ lattice gauge theory in 2D whose Gauss-law operator and other terms are shown in Fig.~\ref{fig:2d}(a). It possesses a dual (magnetic) 1-form $\bbz_2^{B, (1)}$ symmetry generated by
\be 
V_B= \prod_{e \subset \Gamma} X_e,
\ee 
where $\Gamma$ is a closed loop on the dual lattice, i.e., the lattice of $\sigma$ spins. The charge of the 1-form symmetry is a Wilson loop
\be 
W = \prod_{e \subset C } Z_e,
\ee 
where $C$ is a closed loop on the primal lattice.

\subsubsection{Bosonic gauging of $\bbz_2^{B, (1)}$ in bosonic Ising-dual gauge theory}
We can reverse the process by gauging the $\bbz_2^{B, (1)}$ 1-form symmetry  of the Ising-dual $\bbz_2$ gauge theory. To do that, we insert an Ising spin or a qubit for each plaquette $r$ and impose the new Gauss law
\be 
X_e \sigma^z_{r^e_1} \sigma^z_{r^e_2} =1.
\ee 
The original Gauss law in Eq.~(\ref{eq:gaugingGauss}) is then automatically satisfied.
The minimally coupled Hamiltonian takes the form of 
\be 
\tilde{H}'_B = - J \sum_e X_e - g \sum_r \sigma^x_r\prod_{e \subset \partial r} Z_e,
\ee 
with a trivial flatness condition. We then apply the same unitary transformation using $U_B$ in Eq.~(\ref{eq:gaugingU}). It maps
\be 
X_e \sigma^z_{r^e_1} \sigma^z_{r^e_2} \to X_e =1,
\ee 
and the gauged Hamiltonian to 
\be 
H_B = - J \sum_e \sigma^z_{r^e_1} \sigma^z_{r^e_2}  - g \sum_r \sigma^x_r.
\ee 
This is the original TFIM with the $\bbz_2^B$ symmetry generated by $\prod_r \sigma^x_r$.

\subsubsection{Fermionic gauging of $\bbz_2^B$ in TFIM}
\label{sec_fermionic_gauging_boson}

We now gauge the $\bbz_2^B$ symmetry of the TFIM in Eq.~(\ref{eq:TFIM}) by inserting Majorana fermions. Instead of placing Ising spins on the edges, we place a pair of Majorana fermions, $\gamma^e_1$ and $\gamma^e_2$, on each edge $e$ [red lines in Fig.~\ref{fig:2d}(c)]. We assign respectively the two Majorana fermions to the two spins on the plaquettes sharing the edge, i.e., $\sigma^z_{r_1^e}$ and $\sigma^z_{r_2^e}$. We denote the four Majorana fermions assigned to plaquette $r$ [red plaquette in Fig.~\ref{fig:2d}(c)] collectively by $\{\gamma^e_r\}$. Since the dual square lattice is bipartite, we can refer to the Majorana fermions assigned to one dual sublattice as $\gamma$ fermions and those assigned to the other dual sublattice as $\gamma'$. We now impose the generalized Gauss law 
\be 
\sigma^x_r \prod_{e \subset \partial r} \gamma^e_r =1. 
\label{eq:gauging190}
\ee 
The Gauss-law operators commute with each other since there are four Majorana fermions in the product and there is no overlapping fermion for different $r$. There is an implicit ordering of the Majorana fermions in the product. On a torus, we can always order them such that the following condition is satisfied: 
\be 
\prod_r \sigma_r^x = \prod_e (-1)^{F_e} \equiv  \prod_e i \gamma_e' \gamma_e.
\ee
Since the ordering in the bilinear $i \gamma_e' \gamma_e$ is flexible, this condition does not impose a genuine constraint.

With the above Gauss law, the minimally coupled Hamiltonian can be written as 
\be
\begin{split}
 \tilde{H}_B = &- J \sum_{r} \sigma^z_r (-1)^{F_{e_{r, \hat{x}}}} \sigma^z_{r+\hat{x}} + \sigma^z_r(-1)^{F_{e_{r, \hat{y}}}} \sigma^z_{r+\hat{y}}\\
 &- g \sum_r  \sigma^x_r.   
\end{split}
\ee 
We generalize the flatness condition in Eq.~(\ref{eq:flatness}) to fermions as 
\be 
\prod_{\partial e \supset v} (-1)^{F_e} =1.
\label{eq:gausslaw200}
\ee 
We now perform the disentangling unitary transformation using 
\be 
U_F= \prod_r \left(P_r^+ + P_r^-  \prod_{e \subset \partial r} \gamma^e_r \right),
\ee 
where $P_r^{\pm} = (1 \pm \sigma_r^z)/2$. Note that the factors in $U_F$ are mutually commutative.  The Hamiltonian is mapped to 
\be 
H'_F = - J \sum_{e}  (-1)^{F_e}  - g \sum_r  \prod_{e \subset \partial r} \gamma^e_r,
\label{eq:gauging321}
\ee 
where the transformed Gauss law $\sigma^x_r =1$ has been plugged in. The flatness condition in Eq.~(\ref{eq:gausslaw200}) remains invariant under the transformation and becomes the effective Gauss law. This generalized fermionic gauge theory is a direct analog of the bosonic gauge theory in Eq.~(\ref{eq:z2gaugetheory}), where red (blue) lines in Fig.~\ref{fig:2d}(a) represent Majorana fermions (local fermion parity), and can be viewed as a Majorana stabilizer code \cite{bravyiMajorana2010} when $J =0$.  The same result can also be derived using the bond-algebra formalism~\cite{srednickiHidden1980a, fradkinFermion1980, cobanera2011bond}.  Similar to the bosonic counterpart, it has a dual 1-form symmetry $\bbz_2^{F, (1)}$ generated by 
\be 
V_F = \prod_{e \subset \Gamma} (-1)^{F_e},
\ee 
where $\Gamma$ is a closed loop on the dual lattice. It is the fermion parity operator over a closed loop. Similar to the discussion in the last subsection, the dual 1-form $\bbz_2^{F, (1)}$ symmetry can be gauged to recover the original TFIM.  The fermionic gauging procedure described above generalizes to any graph in which every face has an even number of edges, i.e., to graphs whose duals are Eulerian. A generalization using bond algebras has been discussed in Ref.~\cite{nussinovArbitrary2012}.  

\subsubsection{Gauging $\bbz_2^B$ in TFIM with stacking}
We now show that the fermionic gauging $\bbz_2^B$ of the TFIM is equivalent to diagonal gauging $\bbz_2$ after stacking it with a fermionic system. Similar to the 1D case, we first place gauge spins on the edges and relax the Gauss law in Eq.~(\ref{eq:gauging190}) to
\be 
\sigma^x_r \prod_{e \subset \partial r} \gamma^e_r = \prod_{e \subset \partial r} Z_e
\ee 
and impose the flatness condition 
\be 
\prod_{\partial e \supset v} X_e =1.
\ee 
The minimally coupled Hamiltonian takes the form
\be 
\begin{split}
 \tilde{H}_B = &- J \sum_{r} \sigma^z_r X_{e_{r, \hat{x}}} \sigma^z_{r+\hat{x}} + \sigma^z_r X_{e_{r, \hat{y}}} \sigma^z_{r+\hat{y}} \\
 &- g \sum_r  \sigma^x_r -\kappa \sum_e (-1)^{F_e} X_e,   
\end{split}
\ee 
where $\kappa$ is taken to be large. 
Note that we can view $\prod_{e \subset \partial r} \gamma^e_r$, up to a minus sign, as a local fermion parity on the plaquette $r$ with four Majorana fermions. The flatness condition then removes the ambiguity in the minimal coupling scheme as homological Wilson lines of the gauge field $X$ connecting two fermions are equivalent. Applying unitary transformations, we map the Gauss law to $\sigma^x_r  =1$ and the minimally coupled Hamiltonian to  
\be 
\tilde{H}'_B = - J \sum_{e} X_e - g \sum_r  \prod_{e \subset \partial r} Z_e \prod_{e \subset \partial r} \gamma^e_r    -\kappa \sum_{e} (-1)^{F_e} X_e.
\ee  
$\prod_{\partial e \supset v} X_e =1$ remains invariant.  The Hamiltonian reduces to 
\be 
\tilde{H}''_B = - J \sum_{e} X_e - g \sum_r  \prod_{e \subset \partial r} Z_e \prod_{e \subset \partial r} \gamma^e_r
\label{eq:gauging124}
\ee
with 
$X_e =(-1)^{F_e}$ and $\prod_{\partial e \supset v} X_e =1$, in the limit $\kappa \to \infty$. A unitary transformation using
\be 
U''_F = \prod_e \left[ P^+_e + P^-_e (-1)^{F_e} \right], 
\label{eq:gauging234}
\ee 
with $P_e^{\pm} = (1\pm Z_e)/2$, maps $X_e(-1)^{F_e} \to X_e =1$ and the Hamiltonian to  
\be 
H'_F= - J \sum_{e} (-1)^{F_e} - g \sum_r \prod_{e \subset \partial r} \gamma^e_r,
\ee 
subject to the new Gauss law
\be 
\prod_{\partial e \supset v} (-1)^{F_e} =1.
\ee 
Note that the Hamiltonian in Eq.~(\ref{eq:gauging124}) is the gauged Hamiltonian of the bosonic gauge theory in Eq.~(\ref{eq:z2gaugetheory})  if we impose the Gauss law
\be 
X_e (-1)^{F_e} =1.
\ee 
Thus, we have also derived the duality map from the bosonic gauge theory to the fermionic gauge theory by the fermionic gauging of $\bbz_2^B$ (see  Fig.~\ref{fig:triangle}). Furthermore, it is worth noting that by energetically enforcing the Gauss law and flatness conditions --- through the inclusion of these terms in the stacked Hamiltonian --- the resulting model may be viewed as a generating Hamiltonian, from which distinct dual theories emerge in different parameter regimes.

\subsubsection{Bosonic gauging of $\bbz_2^F$ in fermionic Ising-dual gauge theory}
The above discussion indicates how the bosonic gauge theory can be recovered from the fermionic gauge theory through gauging. Instead of gauging the dual 1-form symmetry, we gauge the (0-form) fermion parity of the fermionic gauge theory as follows.  We introduce gauge spins on the edges, impose the Gauss law 
\be  
\prod_{e \subset \partial r} \gamma^e_r  \prod_{e \subset \partial r} Z_e  =1,
\ee 
and modify the original Gauss law to   \be \prod_{\partial e \supset v} (-1)^{F_e}  =\prod_{\partial e \supset v} X_e .\ee 
As mentioned in the last section, $\prod_{e \subset \partial r} \gamma^e_r$ can be viewed a local parity generator associated with plaquette $r$. The minimally coupled Hamiltonian takes the form
\be 
\tilde{H}_F = - J \sum_{e} (-1)^{F_e}X_e -  g \sum_r \prod_{e \subset \partial r} \gamma^e_r.
\ee
Under the unitary transformation using $U''_F$ in Eq.~(\ref{eq:gauging234}), $\prod_{e \subset \partial r} \gamma^e_r =1$ and the Hamiltonian reduces to
\be 
H'_B = - J \sum_e X_e - g \sum_r \prod_{e \subset \partial r} Z_e,
\ee 
with $\prod_{\partial e \supset v} X_e =1$. This is precisely the Hamiltonian of the bosonic gauge theory, and the duality triangle shown in Fig.~\ref{fig:triangle} is now complete.

An alternative way to understand this duality is to reverse the procedure described in the previous section. We tensor the fermionic gauge theory with qubits on the edges in states $X_e =1$. Applying $U''_F$ in Eq.~(\ref{eq:gauging234}) recovers the Hamiltonian in Eq.~(\ref{eq:gauging124}) with the Gauss-law operator $G_e \equiv X_e (-1)^{F_e}$, which also serves as the generator of gauge transformations on the states in the Hilbert space. The gauge can always be fixed such that $\prod_{e \subset \partial r} \gamma^e_r = 1$. To see this,  denote $A_r \equiv  \prod_{e \subset \partial r} Z_e$ and  $B_r \equiv \prod_{e \subset \partial r} \gamma^e_r$. Since $A_r$ and $B_r$ commute, the Hilbert space can be spanned by common eigenstates of both operators. Let $|\Psi\rangle$ be such a state with eigenvalues $\{a_r, b_r\}$. A gauge-invariant state generated from $|\Psi\rangle$ is proportional to $\prod_e (1 +G_e) |\Psi\rangle$. Since $[A_rB_r, G_e] =0$, all states generated by $G_e$ share the same eigenvalues $a_r b_r$ for $A_r B_r$. Therefore, one can always fix the gauge by selecting a state within the orbit generated by $G_e$ to represent $|\Psi\rangle$. In particular, since $B_r$ anticommutes with $G_e$ when $e \subset \partial r$, we can choose a state with $b_r = 1$, i.e., $\prod_{e \subset \partial r} \gamma^e_r = 1$.

\subsection{Gauging and dualities of Majorana lattice and its dual in 2D}
\label{sec_gaugingMaj}
\begin{figure}[tb]
    \centering
    \includegraphics[width=1\linewidth]{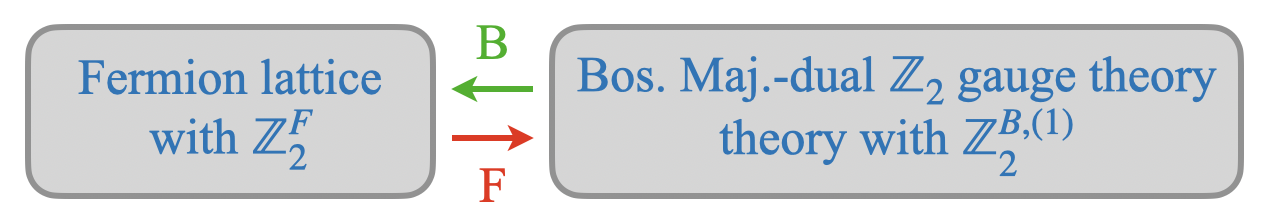}
    \caption{Duality between a 2D fermion lattice with fermion parity $\bbz_2^F$ and its bosonic dual gauge theory with a dual $\bbz_2^{B,(1)}$ 1-form symmetry. Green (red) arrows indicate bosonic (fermionic) gauging. }
    \label{fig:majdual}
\end{figure}

The above discussion of the duality transformations for the TFIM and its dual gauge theories can be extended to the Majorana lattice and its dual theories. Here, the Majorana lattice can be a general fermion lattice that preserves fermion parity $\bbz_2^F$. The relation is shown in Fig.~\ref{fig:majdual}. 

\subsubsection{Bosonic gauging of $\bbz_2^F$ in Majorana lattice}
\label{sec_bos_gauing_Maj}
The general gauging procedure is described in detail in Ref.~\cite{su2025bosonization}, and we briefly outline one possible scheme here. We place a complex fermion or, equivalently, two Majorana fermions on each plaquette $r$ [Fig.~\ref{fig:bos_gauging}(b)]. The local fermion parity is given by $(-1)^{F_r} = i \gamma_r' \gamma_r$. Consider a free Hamiltonian
\be 
H_F = - J \sum_{r} i\gamma_r \gamma'_{r+\hat{x}} + i \gamma_{r+\hat{y}} \gamma'_r - g \sum_r  i \gamma'_r \gamma_r.
\label{eq:gaugingFreeHam}
\ee 
Gauging the total fermion parity $\prod_r (-1)^{F_r} = \prod_r i \gamma_r' \gamma_r$ is similar to the bosonic case. We place gauge spins on the edges of the lattice and impose the Gauss law for each plaquette
\be 
(-1)^{F_r}  \prod_{e \subset \partial r} Z_e  =1,
\ee 
and the flatness condition for each $v$ 
\be 
\prod_{\partial e \supset v} X_e =1. 
\ee 
The gauged Hamiltonian is 
\be 
\tilde{H}_F = - J\sum_{r} i\gamma_r X_{e_{r, \hat{x}}} \gamma'_{r+\hat{x}} + i\gamma_{r+\hat{y}} X_{e_{r, \hat{y}}} \gamma'_r - g \sum_r  i \gamma'_r \gamma_r.
\ee 
The disentangling unitary is given by 
\be 
U_F = \stackrel{\leftarrow}{\prod}_e (P_e^+ + P_e^- S_e),
\label{eq:gaugingU2}
\ee 
where $S_e  = i\gamma_r  \gamma'_{r+\hat{x}}$ for vertical $e = e_{r, \hat{x}}$ and  $S_e = i\gamma'_r \gamma_{r+\hat{y}}$ for horizontal $e = e_{r, \hat{y}}$ as indicated by the cyan arrows in Fig.~\ref{fig:bos_gauging}(b). $P_e^{\pm} = (1\pm Z_e)/2$ are the same projectors as in the bosonic case. Note that since different factors in $U_F$ do not commute, compared to $U_B$ in Eq.~(\ref{eq:gaugingU}), we have implicitly assumed a global ordering of edges in the product. For each plaquette, we order two edges as indicated by the orange numbers in Fig.~\ref{fig:bos_gauging}(b), and then increase to the right and then to the top as indicated by the orange arrows. A factor associated with a smaller index in the product $\stackrel{\leftarrow}{\prod}$ in $U_F$ is placed to the right.   Applying the adjoint action by $U_F$ transforms $(-1)^{F_r}$ to $\prod_{e \subset \partial r} Z_e$ and 
\be 
(-1)^{F_r}  \prod_{e \subset \partial r} Z_e \to (-1)^{F_r} =1.
\ee 
The gauged Hamiltonian becomes
\be 
H_B = - J \sum_{r}  X_{e_{r, \hat{x}}} Z_{e_{r, -\hat{y}}} + X_{e_{r, \hat{y}}} Z_{e_{r, -\hat{x}}} - g \sum_r  \prod_{e \subset \partial r} Z_e,
\label{eq:MajoranaDual}
\ee 
and the flatness condition, which is reinterpreted as the Gauss law, becomes
\be 
\prod_{\partial e \supset v} X_e \prod_{e \supset \partial r_{\text{ne}(v)}} Z_e=1, 
\label{eq:gaugingGauss2}
\ee 
where $r_{\text{ne}(v)}$ denotes the plaquette in the upper right direction of $v$. $H_B$ is the bosonized Hamiltonian of the fermionic $\bbz_2$ gauge theory as in Ref.~\cite{chen2018exact}, and the $C_4$ rotational symmetry of the square lattice is explicitly broken. The terms in the Majorana-dual theory are illustrated in Fig.~\ref{fig:2d}(b), where the Gauss-law operator takes the form $A_v B_p$, with the vertex operator $A_v = \prod_{\partial e \supset v} X_e$ and the plaquette operator $B_r = \prod_{e \in \partial r} Z_e$. The flux-pair creation operators are dressed by additional $Z$ factors to preserve the anticommutation relations of fermion bilinears.  This gauge theory features a dual 1-form symmetry generated by emergent fermionic loops --- (direction-dependent) bosonized products of concatenated fermion bilinears along 1-cycles [see Fig.~\ref{fig:maps}(e, f)]:
\be 
V_x = \prod_{r \cap \Gamma_x \neq \emptyset} X_{e_{r, \hat{x}}} Z_{e_{r, \hat{y}}} ,  \quad V_y = \prod_{r \cap \Gamma_y \neq \emptyset} X_{e_{r, \hat{y}}} Z_{e_{r, \hat{x}}} ,
\ee 
where $\Gamma_{x, y}$ are closed loops on the dual lattice along the horizontal and the vertical direction, respectively. The generators can be deformed by the Gauss law.  The charge of the 1-form symmetry  is again a Wilson loop
\be 
W = \prod_{e \subset C } Z_e,
\ee 
where $C$ is a closed loop on the primal lattice. It is sometimes claimed that the 1-form symmetry in the fermionic $\bbz_2$ gauge theory is anomalous and cannot be coupled to a background field consistently \cite{gaiotto2016spin}. This is because the generators are loops of emergent fermions and cannot condense alone. Nevertheless, we can apply the fermionic gauging procedure to the system to reconstruct the original fermionic system.

\subsubsection{Fermionic gauging of $\bbz_2^{B, (1)}$ in Majorana-dual gauge theory}
To reverse the bosonization process, we insert a pair of Majorana fermions at the center of each plaquette $r$, as shown in Fig.~\ref{fig:2d}(d), and impose the new Gauss law
\be 
\begin{split}
  \left(\prod_{e \subset \partial r} Z_e \right)  i\gamma_r \gamma'_{r +\hat{x}}  \left(X_{e_{r, \hat{x}}} Z_{e_{r, -\hat{y}}} \right)  (-1)^{F_{r+\hat{x}}}  =1, \\
  \left( \prod_{e \subset \partial (r +\hat{y})} Z_e  \right) i \gamma_{r +\hat{y}} \gamma'_r \left(  X_{e_{r, \hat{y}}} Z_{e_{r, -\hat{x}}} \right) (-1)^{F_{r}}=1.
  \label{eq:gauging1form}   
\end{split}
\ee 
These fermion bilinears are reintroduced so that the 1-form generators and the Gauss-law operators become trivial. They are mutually commutative. 
The minimally coupled Hamiltonian is
\be 
\begin{split}
 \tilde{H}_B =& - J \sum_{r}  (-1)^{F_{r+\hat{x}}}  X_{e_{r, \hat{x}}} Z_{e_{r, -\hat{y}}} +  (-1)^{F_r} X_{e_{r, \hat{y}}} Z_{e_{r, -\hat{x}}} \\
 &- g \sum_r (-1)^{F_r} \prod_{e \subset \partial r} Z_e.   
\end{split}
\ee 
We can simplify the expressions by applying a unitary transformation 
\be 
U'_B = \prod_r ( {\cal{P}}_r^+ +{\cal{P}}_r^- Z_{e_{r, -\hat{x}}} Z_{e_{r, \hat{y}}}),
\label{eq:unitary12}
\ee 
where ${\cal{P}}_r^{\pm} = [1 \pm (-1)^{F_r}]/2$. Then the new Gauss law becomes
\be 
i\gamma_r \gamma'_{r +\hat{x}}  \left(X_{e_{r, \hat{x}}} Z_{e_{r+\hat{x}, \hat{y}}} \right)    =1, 
 \quad    i \gamma_{r +\hat{y}} \gamma'_r \left(  X_{e_{r, \hat{y}}} Z_{e_{r+\hat{y}, \hat{x}}} \right)     =1,    
\label{eq:gaugingGauss3}
\ee 
and the Hamiltonian becomes 
\be 
\begin{split}
\tilde{H}'_B = &- J \sum_{r}     X_{e_{r, \hat{x}}} Z_{e_{r, -\hat{y}}} +  X_{e_{r, \hat{y}}} Z_{e_{r, -\hat{x}}} \\
&- g \sum_r (-1)^{F_r} \prod_{e \subset \partial r} Z_e.  
\end{split}
\ee 
Note that the original Gauss law in Eq.~(\ref{eq:gaugingGauss2}) for each $v$ is also transformed into a product of those in Eq.~(\ref{eq:gaugingGauss3}) and is guaranteed to be satisfied. We now apply the inverse of the unitary $U_F$ in Eq.~(\ref{eq:gaugingU2}) by reversing the order of the factors in the product and obtain  
\be 
\begin{split}
X_{e_{r, \hat{x}}} Z_{e_{r, -\hat{y}}} &\to  i\gamma_r \gamma'_{r +\hat{x}} X_{e_{r, \hat{x}}}, \\  X_{e_{r, \hat{y}}} Z_{e_{r, -\hat{x}}} &\to i\gamma'_r \gamma_{r +\hat{y}} X_{e_{r, \hat{y}}},\\ 
(-1)^{F_r} &\to (-1)^{F_r} \prod_{e \subset \partial r} Z_e,
 \end{split}
\ee 
and 
\be 
\begin{split}
i\gamma_r \gamma'_{r +\hat{x}}  \left(X_{e_{r, \hat{x}}} Z_{e_{r+\hat{x}, \hat{y}}} \right)  \to   X_{e_{r, \hat{x}}}   =1, \\
i \gamma_{r +\hat{y}} \gamma'_r \left(  X_{e_{r, \hat{y}}} Z_{e_{r+\hat{y}, \hat{x}}} \right)  \to  X_{e_{r, \hat{y}}}  =1.    
\end{split}
\ee 
Consequently, the Hamiltonian is mapped back to 
\be 
H_F = - J\sum_{r} i\gamma_r \gamma'_{r+\hat{x}} + i \gamma_{r+\hat{y}} \gamma'_r - g \sum_r  i \gamma'_r \gamma_r,
\ee 
the same Hamiltonian in Eq.~(\ref{eq:gaugingFreeHam}) with fermion parity as the dual symmetry. 

\subsubsection{Gauging $\bbz_2^{B, (1)}$ in Majorana-dual gauge theory with stacking}
Similar to the 1D case and the Ising-dual case, the fermionic gauging is equivalent to gauging with stacking. We relax the Gauss law in Eq.~(\ref{eq:gauging1form}) to
\be  
\begin{split}
 \left(\prod_{e \subset \partial r} Z_e \right)  i\gamma_r \gamma'_{r +\hat{x}}  \left(X_{e_{r, \hat{x}}} Z_{e_{r, -\hat{y}}} \right)  (-1)^{F_{r+\hat{x}}}  =\sigma_r^z \sigma_{r+\hat{x}}^z,  \\
\left( \prod_{e \subset \partial (r +\hat{y})} Z_e  \right) i \gamma_{r +\hat{y}} \gamma'_r \left(  X_{e_{r, \hat{y}}} Z_{e_{r, -\hat{x}}} \right) (-1)^{F_{r}}  =\sigma_{r+\hat{y}}^z \sigma_r^z.   
\end{split}   
\ee 
The minimally coupled Hamiltonian is 
\be  
\begin{split}
  \tilde{H}_B =&    - J \sum_{r}  \sigma_{r+\hat{x}}^x  X_{e_{r, \hat{x}}} Z_{e_{r, -\hat{y}}} +   \sigma_r^x X_{e_{r, \hat{y}}} Z_{e_{r, -\hat{x}}}  \\
  &- g \sum_r \sigma_r^x \prod_{e \subset \partial r} Z_e   - \kappa \sum_r \sigma_r^x (-1)^{F_r},   
\end{split}
\ee 
where $\kappa$ is taken to be large.  We first perform a unitary transformation to map the Gauss law to  
\be 
\begin{split}
i\gamma_r \gamma'_{r +\hat{x}}  \left(X_{e_{r, \hat{x}}} Z_{e_{r+\hat{x}, \hat{y}}} \right)    =\sigma_r^z \sigma_{r+\hat{x}}^z, \\ 
i \gamma_{r +\hat{y}} \gamma'_r \left(  X_{e_{r, \hat{y}}} Z_{e_{r+\hat{y}, \hat{x}}} \right)     =\sigma_{r+\hat{y}}^z \sigma_r^z. 
\end{split}
\ee 
Then the Hamiltonian becomes 
\be  
\begin{split}
 \tilde{H}'_B =&  - J \sum_{r}     X_{e_{r, \hat{x}}} Z_{e_{r, -\hat{y}}} +  X_{e_{r, \hat{y}}} Z_{e_{r, -\hat{x}}}\\& - g \sum_r \sigma_r^x \prod_{e \subset \partial r} Z_e   - \kappa \sum_r \sigma_r^x (-1)^{F_r}.
\end{split}
\ee  
The inverse of the unitary $U_F$ in Eq.~(\ref{eq:gaugingU2})  maps them to 
\be 
X_{e_{r, \hat{x}}}   =\sigma_r^z \sigma_{r+\hat{x}}^z, \quad    X_{e_{r, \hat{y}}}     =\sigma_{r+\hat{y}}^z \sigma_r^z, 
\ee 
and
\be  
\begin{split}
 \tilde{H}''_B =&  - J\sum_{r}    i\gamma_r \gamma'_{r +\hat{x}} X_{e_{r, \hat{x}}}  +  i\gamma'_r \gamma_{r +\hat{y}} X_{e_{r, \hat{y}}} \\
 & - g \sum_r \sigma_r^x \prod_{e \subset \partial r} Z_e   - \kappa \sum_r i \gamma'_r \gamma_r \sigma_r^x  \prod_{e \subset \partial r} Z_e.
\end{split}
\ee
Perform another unitary transformation that maps 
\be 
\begin{split}
  \sigma_r^z \sigma_{r+\hat{x}}^z X_{e_{r, \hat{x}}} \to X_{e_{r, \hat{x}}} =1, \\       \sigma_{r+\hat{y}}^z \sigma_r^z  X_{e_{r, \hat{y}}}  \to  X_{e_{r, \hat{y}}} =1,     
\end{split}
\ee 
and we obtain 
\be
\begin{split}
 \tilde{H}'''_B = & - J \sum_{r}    i\gamma_r \gamma'_{r +\hat{x}}  \sigma_r^z \sigma_{r+\hat{x}}^z  +  i\gamma'_r \gamma_{r +\hat{y}}          \sigma_{r+\hat{y}}^z \sigma_r^z \\
&- g \sum_r \sigma_r^x     - \kappa \sum_r  i \gamma'_r \gamma_r \sigma_r^x .  
\end{split}
\label{eq:gauging14}
\ee
The last term becomes an effective Gauss law, $\sigma_r^x(-1)^{F_r} =1 $, for the Hamiltonian in the limit $\kappa \to \infty$. Performing a disentangling unitary maps it to the original Majorana lattice   
\be  
H_F =  - J \sum_{r}    i\gamma_r \gamma'_{r +\hat{x}}   +  i\gamma'_r \gamma_{r +\hat{y}}  
- g \sum_r i \gamma'_r \gamma_r.
\label{eq:Ham_free2}
\ee
Note that Eq.~(\ref{eq:gauging14}) should not be regarded as a gauged Hamiltonian of the TFIM with the Gauss law $\sigma_r^x(-1)^{F_r} =1 $. While $\sigma_r^z \sigma_{r+\hat{x}}^z$ and $ \sigma_{r+\hat{y}}^z \sigma_r^z $ always commute, the corresponding fermionic terms $i\gamma_r \gamma'_{r +\hat{x}}$ and $i\gamma'_r \gamma_{r +\hat{y}}$ do not. Nevertheless, the last step can be reversed and viewed as an alternative way to gauge the fermion parity, compared with the approach in Sec.~\ref{sec_bos_gauing_Maj}.

\subsubsection{Fermionic gauging of $\bbz_2^F$ in Majorana lattice}
We briefly outline the fermionic gauging of the Majorana Hamiltonian above by generalizing the discussions in Secs.~\ref{sec_fermionic_analog} and \ref{sec_fermionic_gauging_boson}. Majorana fermions $\tilde{\gamma}$ are placed on the edges, and we impose the Gauss-law
\be 
(-1)^{F_r}  \prod_{e \subset \partial r} \tilde{\gamma}^e_r =1,
\ee 
together with the flatness condition
\be 
\prod_{\partial e \supset v} (-1)^{F_e} =1.
\ee 
The notation follows that of Sec.~\ref{sec_fermionic_gauging_boson}. To disentangle the system, a nonlocal unitary transformation is required,
\be 
U'_F = \prod_r ( \tilde{{\cal{P}}}_r^+ +\tilde{{\cal{P}}}_r^- \gamma'_r), 
\ee 
where $\tilde{{\cal{P}}}_r^{\pm} = (1 \pm \prod_{e \subset \partial r} \tilde{\gamma}^e_r)/2$. As in  Sec.~\ref{sec_fermionic_analog}, an ordering of the factors is assumed. Under this transformation, one of the first two terms in Eq.~(\ref{eq:Ham_free2}) is mapped to a nonlocal operator dressed by plaquette operators $\prod_{e \subset \partial r} \tilde{\gamma}^e_r$. The resulting fermoinic gauge theory has a dual 1-form symmetry $\bbz_2^{F, (1)}$ generated by $V_F = \prod_{e \subset \Gamma} (-1)^{F_e}$, where $\Gamma$ is a closed loop on the dual lattice. This 1-form symmetry can be gauged, in a nonlocal manner, by reintroducing the $\gamma'$ fermions.

\section{Mapping to ancilla-decorated Ising models}
\label{sec_mapping}
\begin{figure}[th]
    \centering
    \includegraphics[width=0.88\linewidth]{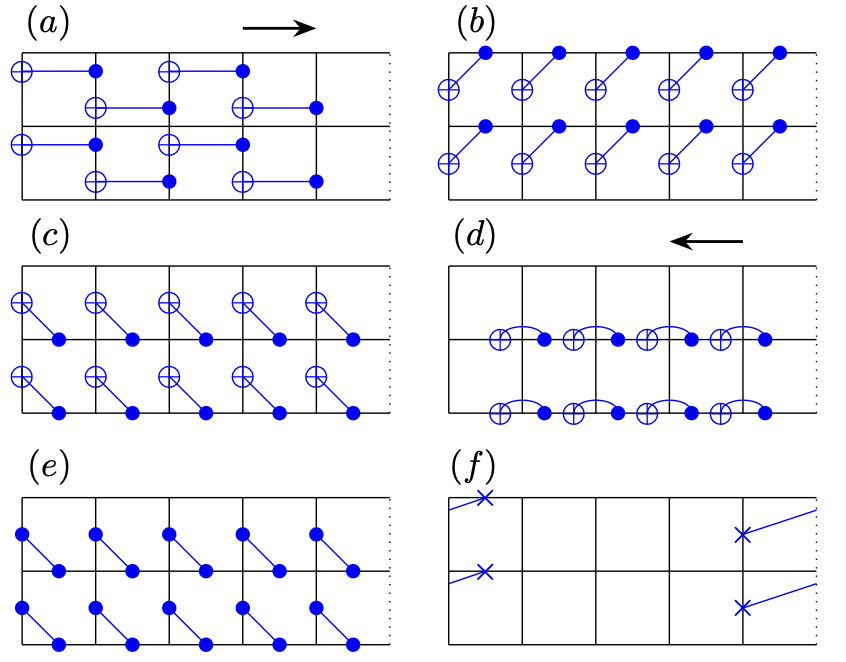}
    \caption{Quantum circuits used to map the gauge theories to ancilla-decorated Ising models. The rightmost dotted edges are identified with the leftmost edges. (a-d) CNOT gates are applied. In (a) and (d), the arrows indicate the sequential application of the CNOT cascades. (e) The CZ gate layer is required for the Majorana-dual theory.  (f) The SWAP gate layer is applied last to the Ising-dual gauge theory across the boundary, while additional layers for the Majorana-dual gauge theory are shown in Fig.~\ref{fig:circuits2}.}
  \label{fig:circuit}
\end{figure}

\begin{figure}[bh]
    \centering
    \includegraphics[width=0.88\linewidth]{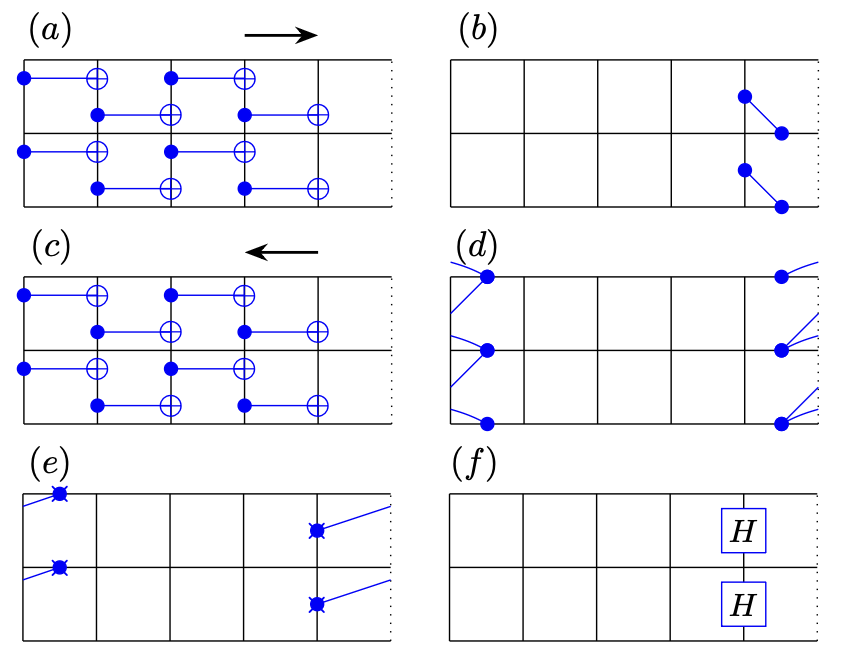}
     \caption{Additional circuit layers to Fig.~\ref{fig:circuit} applied to the bosonic Majorana-dual $\bbz_2$ gauge theory on a torus. The rightmost dotted edges are identified with the leftmost edges. (a, c) The arrows indicate the sequential application of the CNOT cascades. (b, d) SWAP gates are applied.  (e) The gates are fermionic SWAP gates defined as fSWAP = CZ $\cdot$ SWAP. (f) Hadamard gates are applied to the last column of vertical edges.}
  \label{fig:circuits2}
\end{figure} 

\begin{figure*}
    \centering
    \includegraphics[width=0.8\linewidth]{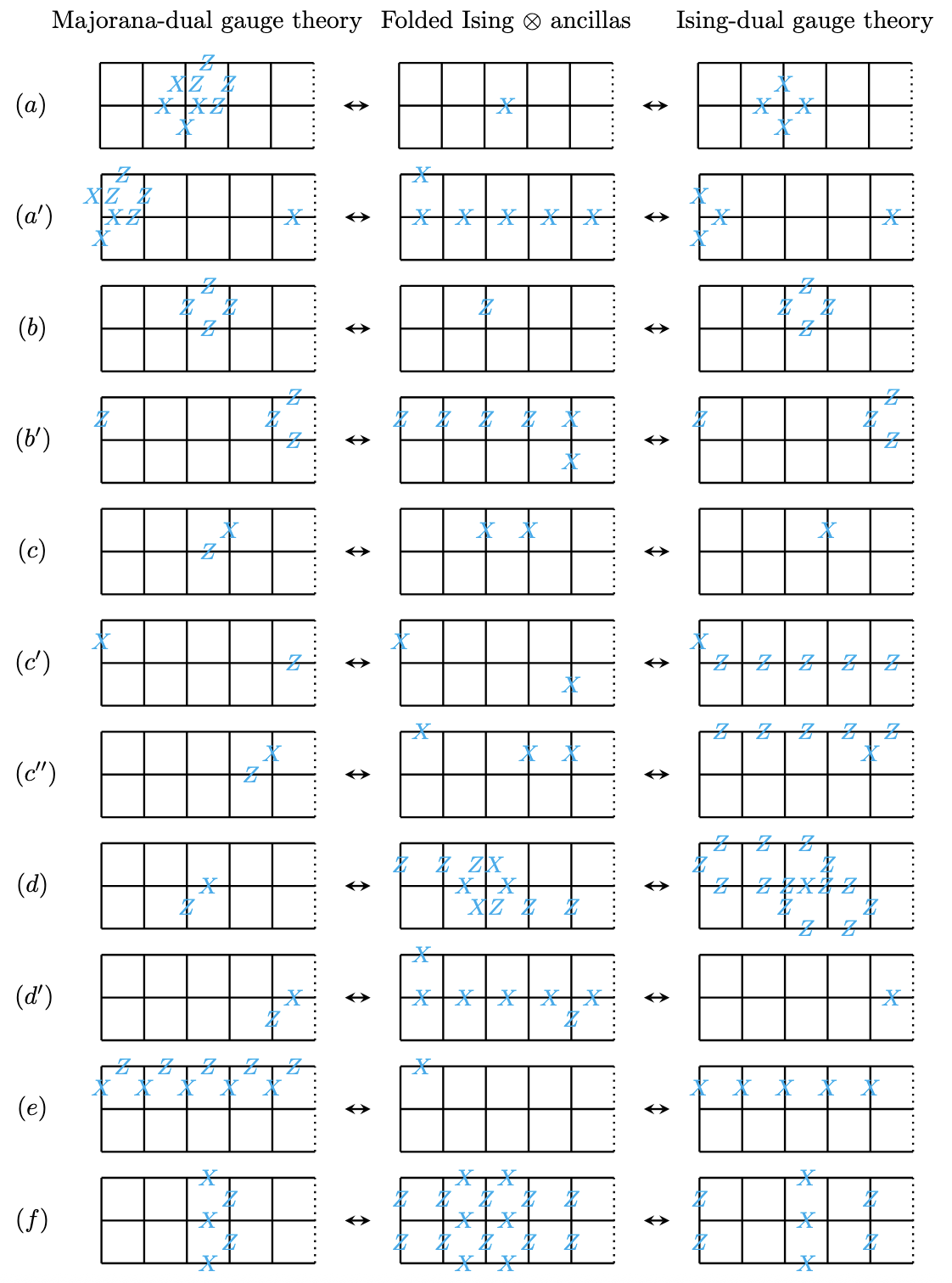}
    \caption{Operator mappings of the Majorana-dual gauge theory (left) to the ancilla-decorated Ising model (center) and to the Ising-dual gauge theory (right). (a, a$'$) Gauss-law operators. (b, b$'$) Plaquette operators. (c, c$'$, c$''$) Dressed $X$ operators on vertical edges. (d, d$'$) Dressed $X$ operators on horizontal edges. (e) Horizontal 1-form symmetry generator. (f) Vertical 1-form symmetry generator. The rightmost dotted edges of each lattice are identified with the leftmost edges.}
\label{fig:maps}
\end{figure*} 

In the following, we focus on bosonic gauge theories. We relate the Majorana-dual gauge theory on a square lattice to the Ising-dual gauge theory by mapping both to ancilla-decorated Ising models. It is known that coupling a system to a flat (bosonic) gauge field corresponds to summing over different twisted sectors \cite{gaiotto2015generalized}. When boundary effects are neglected by taking the system to be infinite, the Gauss-law operators can be transformed to act on a decoupled subsystem. To be more concrete, consider a sequential circuit $U$ composed of the four layers of local unitary gates shown in Fig.~\ref{fig:circuit} (a-d), where each layer consists of controlled-$X$ (CNOT) gates (see Appendix~\ref{sec_gates}). The ordering of the CNOT cascades in (a) and (d) is indicated by the arrows above. The adjoint action of $U$ on the Ising-dual gauge theory maps the Gauss-law operator $G_v =1$ to $X_e =1$ on a horizontal edge, and thereby effectively decoupling the Gauss law constraints from the spins on the vertical edges. One can verify that, up to boundary terms, the Ising-dual gauge theory is mapped to the TFIM on the vertical edges. Adding the fifth layer of controlled-$Z$ (CZ) gates, as shown in Fig.~\ref{fig:circuit}(e), similarly decouples the modified Gauss law of the Majorana-dual gauge theory [in Fig.~\ref{fig:2d}(b)] \cite{chen2023equivalence}. We now incorporate periodic boundary conditions on a torus. In this case, the CNOT gates are not allowed to act across the periodic boundary on the right in Fig.~\ref{fig:circuit}(a, d). Additionally, we perform extra swap operations as indicated in Fig.~\ref{fig:circuit}(f) for the Ising-dual gauge theory. The Gauss-law operators across the periodic boundary are reduced to Ising couplings $X_{e} X_{e +\hat{y}}$, where $e$ and $e+\hat{y}$ are neighboring horizontal edges in the first column [see Fig.~\ref{fig:maps}(a$'$)]. Note that, consistent with the global constraint $\prod_v G_v =1$, the ferromagnetic Ising couplings in the first column introduces a two-fold degeneracy that contributes to the four-fold degeneracy of the toric code. The Gauss-law operators across the periodic boundary in the Majorana-dual gauge theory can be mapped to the same configuration [see Figs.~\ref{fig:circuits2} and \ref{fig:maps}(a$'$)].

After these operations, the qubits/spins on the horizontal edges become decoupled from those on the vertical edges. The former can thus be viewed as ancilla qubits associated with each qubit on a vertical edge. The transformed Gauss laws require the ancillas to be in states corresponding to the ground state of a parent Hamiltonian, while excitations in the ancillas encode the charges associated with the Gauss laws. Under the unitary transformation, the horizontal 1-form symmetry generator of the Majorana-dual gauge theory, shown in Fig.~\ref{fig:maps}(e), is mapped to a single $X_e$ acting on the first column, whereas the vertical generator becomes the global spin-flip acting on the vertical edges after decoupling.

\section{Anyon mapping and transmutation}
\label{sec_transmutation}
The unitary equivalences between the bosonic gauge theories and the ancilla-decorated Ising models can be composed into a direct map between the gauge theories (Fig.~\ref{fig:maps}). Under this map, the Gauss-law operators are mapped onto each other, and the plaquette operators remain invariant. Consequently, the fermionic toric code is unitarily equivalent to the bosonic one. However, since the unitary transformation is implemented by a linear-depth local circuit, the induced anyon mappings are nontrivial. In the Ising-dual gauge theory, an $X$ creates a pair of fluxes, i.e., $m$ anyons, along either the horizontal or vertical direction. In the Majorana-dual theory, this role is played by a $Z$-decorated $X$. As shown in Fig.~\ref{fig:maps}, under the unitary map, a $Z$-decorated $X$ is mapped to a bare $X$, if the boundary terms are neglected. In contrast, in the vertical direction, a $Z$-decorated $X$ in the Majorana-dual theory is mapped to an $X$ operator dressed by a large Wilson loop extending to the boundary. Conversely, in the vertical direction, an $X$ operator in the Ising-dual gauge theory is mapped to a $Z$-decorated $X$ similarly attached to a large Wilson loop. This direction-dependent flux attachment accounts for the anyon transmutation between the two theories. 

\begin{figure}[tb]
    \centering
    \includegraphics[width=0.95\linewidth]{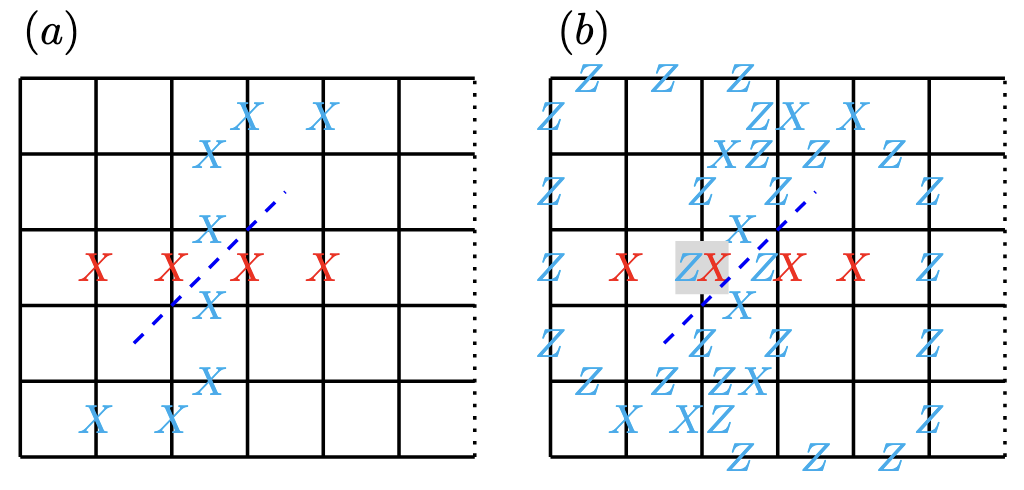}
    \caption{Exchange statistics and direction-dependent transmutation of flux excitations. (a) Two pairs of fluxes in the bosonic toric code are created by horizontal string operators. One string (cyan) is deformed to cross the other (red). At the intersection, they split along the dashed diagonal into two new strings, which are then deformed back to the horizontal direction. (b) Similar process in the fermionic toric code, except that deformation along the vertical direction is accompanied by Wilson loops extending to the boundary. The anticommutation between $X$ and $Z$ on the shaded edge introduces a minus sign, accounting for the fermionic statistics of the flux excitations.}
\label{fig:braiding}  
\end{figure}

We can verify the braiding statistics as illustrated in Fig.~\ref{fig:braiding}. In the Ising-dual gauge theory, two pairs of fluxes are created by two 't Hooft lines along the edges of the dual lattice. As shown in Fig.~\ref{fig:braiding}(a), one 't Hooft line (cyan) can be deformed to cross another (red). At the intersecting square, the two line operators can split along the dashed diagonal without introducing any sign. The operators can then be  deformed back to the original configuration, effectively exchanging the positions of two fluxes. The absence of a sign change indicates that the fluxes are bosonic. After the directional flux-attachment, we can repeat the same exchange operations. The corresponding crossing configuration is shown in Fig.~\ref{fig:braiding}(b). When the two operators are split in the overlapping region, the anticommutation of the $X$ and $Z$ operator in the shaded square introduces a minus sign, revealing the fermionic statistics of the flux excitations in the fermionic toric code. We can show that this fermionic character can be traced back to a folded Majorana chain.

\section{Jordan--Wigner and Kramers--Wannier}
\label{sec_jw}
\begin{figure}[tb]
    \centering
    \includegraphics[width=0.65\linewidth]{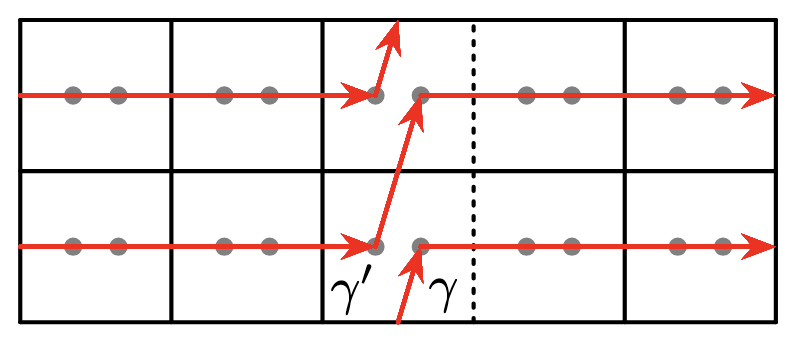}
    \caption{Folded path used in the JW transformation on a torus. The periodic boundary (dotted line) is shifted to the center for clarity.}
\label{fig:maj}
\end{figure}

After decoupling the ancillas on the horizontal edges by fixing $X_e =1$ (including those in the first column), the Majorana-dual gauge theory reduces to an Ising model on the vertical edges. This Ising model is precisely a folded Ising chain derived from a folded Majorana chain via the JW transformation. The folded path on a torus shown in Fig.~\ref{fig:maj} covers the 2D space. Each plaquette hosts two Majorana fermions, $\gamma$ and $\gamma'$, where the $\gamma'$ fermion on the boundary plaquette is connected to the $\gamma$ fermion on the plaquette one row above. Along the folded path, there are two types of spin operators, $Z$ and $\prod_{j<k} Y_j Z_{j+1}\cdots Z_{k-1} Y_k$ (including $X_j X_{j+1}$), which are exactly the JW images of fermion bilinears. Bilinears connecting different rows are mapped to string operators, and the bilinear of the two fermions within the boundary plaquette is likewise nonlocal with respect to the Majorana chain. This subtlety is reflected in Fig.~\ref{fig:maps}(b$'$, d$'$) where the corresponding spin operators for the plaquette and the hopping terms are interchanged. As discussed above, these string operators represent the creation and annihilation of emergent fermions. In two dimensions, their fermionic statistics manifest as the sign change under the exchange operation described earlier.    

Setting $X_e =-1$ on the horizontal edges of the first column introduces a twist in the Ising coupling for each row, as shown in Fig.~\ref{fig:maps}(c$''$). As noted above, these $X_e$ operators correspond to the 1-form symmetry generators along the horizontal direction in the Majorana-dual gauge theory. Consequently, within the eigenspace of the 1-form symmetry with eigenvalue $+1$, the mapping to the folded Ising chain remains one-to-one. 

The KW (self-)dualities constructed in Ref.~\cite{su2025bosonization} for the Majorana-dual gauge theories originate from translations of the Majorana lattice. These self-dualities are not obtained by gauging a 1-form symmetry, since the corresponding dual symmetry is a 0-form symmetry. Because the 1-form symmetry along the vertical direction is mapped to the $\bbz_2^B$ spin-flip symmetry of the Ising model defined on the vertical edges [Fig.~\ref{fig:maps}(f)],  these KW (self-)dualities should be compared with those obtained by gauging the $\bbz_2^B$ symmetry of the folded Ising chain. When the horizontal boundary is neglected, the mappings of the bulk operators coincide, as the global $\bbz_2$ symmetry effectively decomposes into subsystem (or gauge-like) symmetries \cite{nussinovsymmetry2009} acting independently on different rows. Once the toroidal geometry is taken into account, however, the $\bbz_2^B$ symmetry of the folded Majorana chain cannot be interpreted as a subsystem symmetry.
Bosonic gauging $\bbz_2^B$ of the Ising model shifts operators across the boundary to the adjacent row, corresponding to a minimal translation of the folded chain. Consequently, the KW (self-)dualities in Ref.~\cite{su2025bosonization}, derived from the foliated translation of the Ising chain, are not generated by gauging the $\bbz_2^B$ symmetry. Thus, these  KW (self-)dualities are distinct from those obtained from  gauging higher-form symmetry or subsystem symmetries.

\section{Generalization to 3D}
\label{sec_3d}
The analysis above extends naturally to higher dimensions. Dual theories of the 3D TFIM can be obtained using bosonic gauging and fermionic gauging. The bosonic Majorana-dual theory can be derived by gauging the fermion parity following the general procedure of Ref.~\cite{su2025bosonization} with a chosen ordering of faces (or dual edges), or equivalently using operator-algebraic mappings~\cite{chen2019bosonization}. On the cubic lattice with 3D torus geometry, the Gauss-law operators in the bosonic Ising-dual theory involve only the plaquette operators on the faces, whereas in the Majorana-dual theory they are dressed due to the anticommutation relations of fermions (see Appendix~\ref{sec:3D}). As in 2D, both bosonic gauge theories can be mapped to ancilla-decorated Ising models, where the transformed Gauss-law operators act solely on edges lying in planes normal to a chosen axis and thereby decouple from the degrees of freedom along that axis. On a 3D torus, one first constructs a local circuit that decouples the bulk, and then the remaining steps needed to decouple the boundary terms are essentially the same as those of the 2D case. The directional transmutation and the correspondence with the JW transformation along a lattice-covering folded path and the KW self-dualities are natural generalizations of the 2D scenario.

\section{Conclusion}
\label{sec_conclusion}
In this work, we generalized the conventional bosonic $\bbz_2$ gauging procedure to fermionic gauging, providing a unified framework for fermionic and bosonic lattice systems (Fig.~\ref{fig:relation}). We constructed local unitary circuits connecting gauge theories dual to the TFIM and to the free Majorana lattice with ancilla-decorated Ising models, revealing direction-dependent anyon transmutations. Furthermore, we have shown that the KW self-dualities are closely related to, yet distinct from, those obtained by gauging.

Several directions merit further exploration. The parallel structure of bosonic and fermionic gauging and dualities aligns naturally with the symmetry topological field theory framework~\cite{wenTopological2024, huangFermionic2025, bhardwajFermionic2025}, and calls for systematic development on lattices and in higher dimensions. The relation of this framework to mappings between bosonic and fermionic topological phases also warrants investigation. The direction-dependent transmutations identified here remain largely unexplored. The demonstrated unitary equivalences between distinct systems could enable exotic state preparation and quantum simulation of fermionic models via measurement and feedforward. It would also be interesting to examine potential implications for finite-temperature topological orders~\cite{zhou2025finite} and single-shot error correction~\cite{leeChiral2025} in fermionic toric codes. Finally, one might even envision a unified gauging procedure incorporating supersymmetry or superconnections \cite{quillenSuperconnections1985}.

\section*{Acknowledgements}
This work is dedicated to C. N. Yang. The author thanks Aashish Clerk and Ivar Martin for their valuable support. This research was partially supported by the Simons Foundation through a Simons Investigator Award (Grant No. 669487) and by the US Department of Energy, Office of Science, Basic Energy Sciences, Materials Sciences and Engineering Division.

\begin{appendix}
\numberwithin{equation}{section}

\section{Jordan--Wigner transformation}
\label{app_jw}
Complex fermions, $c_j$ and $c^{\dagger}_j$, on a 1D chain satisfy the anticommutation relations
\be 
\{c_j, c^{\dagger}_{k}\} = \delta_{jk}, \quad \{c_j, c_{k}\} =0, \quad \{c_j^{\dagger}, c^{\dagger}_{k}\} =0.
\ee 
The Majorana fermions, $\gamma'_j$ and $\gamma_j$, are defined by
\be 
\gamma'_{j} = c_j + c_j^{\dagger}, \quad \gamma_{j} = i(c_j^{\dagger} - c_j).
\ee 
They satisfy the relations
\be 
\{\gamma_{j}, \gamma_{k}\} = 2\delta_{jk}, \quad \{\gamma'_{j}, \gamma'_{k}\} = 2\delta_{jk}, \quad \{\gamma_{j}, \gamma'_{k}\} =0.
\ee
The JW transformation expresses
\be 
\gamma_j = \left( \prod_{k<j} Z_k \right) (-Y_j), \quad \gamma'_j = \left( \prod_{k<j} Z_k \right) X_j.
\label{eq:JW}
\ee 
In this convention, 
\be 
i\gamma'_j \gamma_j =Z_j , \quad 
i \gamma_j \gamma'_{j+1} =  X_j X_{j+1}, \quad   i \gamma_{j+1} \gamma'_{j} = Y_j Y_{j+1}.
\ee 

The spin operators in the Majorana-dual gauge theory on the square lattice shown in Fig.~\ref{fig:2d}(b) correspond to
\be 
\begin{split}
    \prod_{e \subset \partial r} Z_e &\to i \gamma'_r \gamma_r, \\  X_{e_{r, \hat{x}}} Z_{e_{r, -\hat{y}}} &\to i \gamma_{r} \gamma'_{r+\hat{x}} , \\ X_{e_{r, \hat{y}}} Z_{e_{r, -\hat{x}}}  &\to i \gamma_{r+\hat{y}} \gamma'_r.
\end{split}
\ee 
Under the unitary transformations discussed in Sec.~\ref{sec_mapping}, the Gauss-law operator becomes decoupled, and these spin operators are mapped to those in the Ising model, as illustrated in Fig.~\ref{fig:maps}. Along the folded path shown in Fig.~\ref{fig:maj}, the operators in the Ising model correspond to those obtained via the JW transformation of the fermion bilinears, up to a half-plaquette shift. A minor subtlety arises for the mappings on the rightmost plaquettes shown in Fig.~\ref{fig:maps}(b$'$, d$'$): the ordering of the fermion bilinears is reversed. This is corrected by removing the minus sign in front of $Y_j$ for $\gamma_j$ in Eq.~(\ref{eq:JW}) on these plaquettes.

\section{Gate actions}
\label{sec_gates}
The actions of the gates used in this work are defined as follows.
\begin{itemize}
    \item CZ gate
    \begin{subequations}
    \begin{equation*}
     \text{CZ} (X \otimes \1) \text{CZ} = X \otimes Z, \quad \text{CZ} (\1 \otimes X) \text{CZ} = Z \otimes X,  \\ 
    \end{equation*} 
    \begin{equation*}
        \text{CZ} (Z \otimes \1) \text{CZ} = Z \otimes \1, \quad \text{CZ} (\1 \otimes Z) \text{CZ} = \1 \otimes Z. 
    \end{equation*}
    \end{subequations}
    \item CNOT or CX gate
    \begin{subequations}
     \begin{equation*}
    \text{CX} (X \otimes \1) \text{CX} = X \otimes X, \quad \text{CX} (\1 \otimes X) \text{CX} = \1 \otimes X,
    \end{equation*} 
     \begin{equation*}
    \text{CX} (Z \otimes \1) \text{CX} = Z \otimes \1, \quad \text{CX} (\1 \otimes Z) \text{CX} = Z \otimes Z. 
    \end{equation*}
    \end{subequations}
    \item SWAP gate
    \begin{subequations}
    \begin{align*}
        \text{SWAP} (X \otimes \1) \text{SWAP} &= \1 \otimes X,\\
          \text{SWAP} (\1 \otimes X) \text{SWAP} &= X \otimes \1,\\
          \text{SWAP} (Z \otimes \1) \text{SWAP} &= \1 \otimes Z, \\ \text{SWAP} (\1 \otimes Z) \text{SWAP} &= Z \otimes \1.
    \end{align*}  
    \end{subequations}
    \item Hadamard gate $H$
    \begin{equation*}
    H X H = Z, \quad H Z H = X.
    \end{equation*} 
\end{itemize}

\section{Bosonic $\bbz_2$ gauge theories on cubic lattice}
\label{sec:3D}
In this appendix, we present more details about the bosonic $\bbz_2$ gauge theories on the cubic lattice with toroidal geometry.

\subsection{Operators in bosonic gauge theories}
\begin{figure}[H]
    \centering
    \includegraphics[width=1\linewidth]{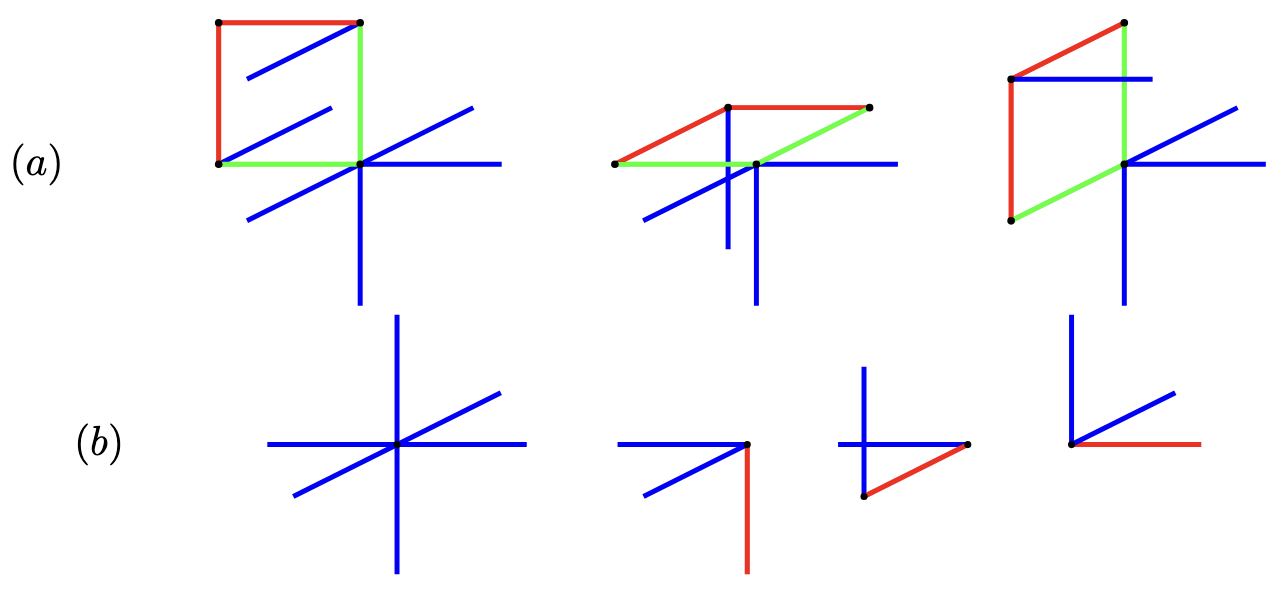}
    \caption{Operators in the Majorana-dual gauge theory on the cubic lattice. (a) Three Gauss-law operators. The overall minus signs are omitted. (b) Vertex operator and dressed $Z$ operators. Pauli operators are indicated by color: $X$ (blue), $Z$ (red), and $Y$ (green).}
    \label{fig:3d}
\end{figure}

The bosonization of the Majorana lattice can be obtained by gauging the fermion parity following the general procedure of Ref.~\cite{su2025bosonization} with a chosen ordering of faces (or dual edges), or equivalently using operator algebraic mappings in Ref.~\cite{chen2019bosonization}. 
When the dual edges of the cubic lattice are ordered from top to bottom, left to right, and back to front, the gauging procedure yields the representation shown in Fig.~\ref{fig:3d}, after a global rotation of basis $X \leftrightarrow Z$. This representation is also used in Ref.~\cite{zhou2025finite}. The first row includes three Gauss-law operators. In the corresponding Ising-dual version, they are simply plaquette operators that are products of four $Z$ operators \cite{dennisTopological2002}. The second row includes the vertex operator and three dressed $Z$ operators. In the Ising-dual gauge theory, the $Z$ operators are undressed. 

Since the product of the three Gauss-law operators associated with the faces of each cube equals the identity, only two  of the Gauss law constraints per cube are independent. These constraints,  sometimes referred to as meta-check conditions, enforce that excitations corresponding to plaquette-operator violations form loops.  

\subsection{Disentangling unitary circuits}
As in the 2D case, we aim to construct a unitary circuit that decouples the Gauss-law operators in both the bosonic Ising-dual and Majorana-dual $\bbz_2$ gauge theories, so that they act only on ancillas. For clarity, we temporarily ignore the rightmost periodic boundary shown in the following figures and map the three types of Gauss-law operators to those acting on edges lying in planes perpendicular to the $x$-axis as follows.     

\begin{subequations}
    \be 
    \begin{tikzpicture}[scale=1.2, line cap=round, line join=round, baseline=(current bounding box.center)]

  \def\Nx{5}       
  \def\Ny{1}       
  \def\Nz{1}       
  \def\angleY{35}  

  \def\a{1.0}

  \pgfmathsetmacro{\bx}{\a}                
  \pgfmathsetmacro{\by}{0}
  \pgfmathsetmacro{\yx}{\a*cos(\angleY)}   
  \pgfmathsetmacro{\yy}{\a*sin(\angleY)}
  \pgfmathsetmacro{\zx}{0}                 
  \pgfmathsetmacro{\zy}{\a}

  \foreach \ix in {0,...,4}{
    \foreach \iy in {0,...,\Ny}{
      \foreach \iz in {0,...,\Nz}{
        \pgfmathsetmacro{\x}{\ix*\bx + \iy*\yx + \iz*\zx}
        \pgfmathsetmacro{\y}{\ix*\by + \iy*\yy + \iz*\zy}

        \ifnum\ix<\Nx
          \pgfmathsetmacro{\xnext}{(\ix+1)*\bx + \iy*\yx + \iz*\zx}
          \pgfmathsetmacro{\ynext}{(\ix+1)*\by + \iy*\yy + \iz*\zy}
          \draw[line width=0.6pt, gray] (\x,\y)--(\xnext,\ynext);
        \fi

        \ifnum\iy<\Ny
          \pgfmathsetmacro{\xnext}{\ix*\bx + (\iy+1)*\yx + \iz*\zx}
          \pgfmathsetmacro{\ynext}{\ix*\by + (\iy+1)*\yy + \iz*\zy}
          \draw[line width=0.6pt, gray] (\x,\y)--(\xnext,\ynext);
        \fi

        \ifnum\iz<\Nz
          \pgfmathsetmacro{\xnext}{\ix*\bx + \iy*\yx + (\iz+1)*\zx}
          \pgfmathsetmacro{\ynext}{\ix*\by + \iy*\yy + (\iz+1)*\zy}
          \draw[line width=0.6pt, gray] (\x,\y)--(\xnext,\ynext);
        \fi

        \filldraw[fill=black] (\x,\y) circle (0.02);
      }
    }
  }

\foreach \ix in {5}{
    \foreach \iy in {0,...,\Ny}{
      \foreach \iz in {0,...,\Nz}{
        \pgfmathsetmacro{\x}{\ix*\bx + \iy*\yx + \iz*\zx}
        \pgfmathsetmacro{\y}{\ix*\by + \iy*\yy + \iz*\zy}

        \ifnum\iy<\Ny
          \pgfmathsetmacro{\xnext}{\ix*\bx + (\iy+1)*\yx + \iz*\zx}
          \pgfmathsetmacro{\ynext}{\ix*\by + (\iy+1)*\yy + \iz*\zy}
          \draw[line width=0.6pt, dotted, gray] (\x,\y)--(\xnext,\ynext);
        \fi

        \ifnum\iz<\Nz
          \pgfmathsetmacro{\xnext}{\ix*\bx + \iy*\yx + (\iz+1)*\zx}
          \pgfmathsetmacro{\ynext}{\ix*\by + \iy*\yy + (\iz+1)*\zy}
          \draw[line width=0.6pt, dotted, gray] (\x,\y)--(\xnext,\ynext);
        \fi

        \filldraw[fill=black] (\x,\y) circle (0.02);
      }
    }
  }

      \pgfmathsetmacro{\xA}{2*\bx + 0.*\yx+ 1.*\zx}
      \pgfmathsetmacro{\yA}{2*\by + 0.*\yy+ 1.*\zy}

      \pgfmathsetmacro{\xAnext}{2*\bx + 1*\yx+ 1.*\zx}
      \pgfmathsetmacro{\yAnext}{2*\by + 1*\yy+ 1.*\zy}

     \draw[line width=1.0pt, red] (\xA,\yA)--(\xAnext,\yAnext);

\end{tikzpicture}
    \ee 
    \be 
    \begin{tikzpicture}[scale=1.2, line cap=round, line join=round, baseline=(current bounding box.center)]

  \def\Nx{5}       
  \def\Ny{1}       
  \def\Nz{1}       
  \def\angleY{35}  

  \def\a{1.0}

  \pgfmathsetmacro{\bx}{\a}                
  \pgfmathsetmacro{\by}{0}
  \pgfmathsetmacro{\yx}{\a*cos(\angleY)}   
  \pgfmathsetmacro{\yy}{\a*sin(\angleY)}
  \pgfmathsetmacro{\zx}{0}                 
  \pgfmathsetmacro{\zy}{\a}

  \foreach \ix in {0,...,4}{
    \foreach \iy in {0,...,\Ny}{
      \foreach \iz in {0,...,\Nz}{
        \pgfmathsetmacro{\x}{\ix*\bx + \iy*\yx + \iz*\zx}
        \pgfmathsetmacro{\y}{\ix*\by + \iy*\yy + \iz*\zy}

        \ifnum\ix<\Nx
          \pgfmathsetmacro{\xnext}{(\ix+1)*\bx + \iy*\yx + \iz*\zx}
          \pgfmathsetmacro{\ynext}{(\ix+1)*\by + \iy*\yy + \iz*\zy}
          \draw[line width=0.6pt, gray] (\x,\y)--(\xnext,\ynext);
        \fi

        \ifnum\iy<\Ny
          \pgfmathsetmacro{\xnext}{\ix*\bx + (\iy+1)*\yx + \iz*\zx}
          \pgfmathsetmacro{\ynext}{\ix*\by + (\iy+1)*\yy + \iz*\zy}
          \draw[line width=0.6pt, gray] (\x,\y)--(\xnext,\ynext);
        \fi

        \ifnum\iz<\Nz
          \pgfmathsetmacro{\xnext}{\ix*\bx + \iy*\yx + (\iz+1)*\zx}
          \pgfmathsetmacro{\ynext}{\ix*\by + \iy*\yy + (\iz+1)*\zy}
          \draw[line width=0.6pt, gray] (\x,\y)--(\xnext,\ynext);
        \fi

        \filldraw[fill=black] (\x,\y) circle (0.02);
      }
    }
  }

\foreach \ix in {5}{
    \foreach \iy in {0,...,\Ny}{
      \foreach \iz in {0,...,\Nz}{
        \pgfmathsetmacro{\x}{\ix*\bx + \iy*\yx + \iz*\zx}
        \pgfmathsetmacro{\y}{\ix*\by + \iy*\yy + \iz*\zy}

        \ifnum\iy<\Ny
          \pgfmathsetmacro{\xnext}{\ix*\bx + (\iy+1)*\yx + \iz*\zx}
          \pgfmathsetmacro{\ynext}{\ix*\by + (\iy+1)*\yy + \iz*\zy}
          \draw[line width=0.6pt, dotted, gray] (\x,\y)--(\xnext,\ynext);
        \fi

        \ifnum\iz<\Nz
          \pgfmathsetmacro{\xnext}{\ix*\bx + \iy*\yx + (\iz+1)*\zx}
          \pgfmathsetmacro{\ynext}{\ix*\by + \iy*\yy + (\iz+1)*\zy}
          \draw[line width=0.6pt, dotted, gray] (\x,\y)--(\xnext,\ynext);
        \fi

        \filldraw[fill=black] (\x,\y) circle (0.02);
      }
    }
  }

      \pgfmathsetmacro{\xA}{2*\bx + 0.*\yx+ 0.*\zx}
      \pgfmathsetmacro{\yA}{2*\by + 0.*\yy+ 0.*\zy}

      \pgfmathsetmacro{\xAnext}{2*\bx + 0*\yx+ 1.*\zx}
      \pgfmathsetmacro{\yAnext}{2*\by + 0*\yy+ 1.*\zy}

     \draw[line width=1.0pt, red] (\xA,\yA)--(\xAnext,\yAnext);

\end{tikzpicture}
    \ee 
    \be
    \begin{tikzpicture}[scale=1.2, line cap=round, line join=round, baseline=(current bounding box.center)]

  \def\Nx{5}       
  \def\Ny{1}       
  \def\Nz{1}       
  \def\angleY{35}  

  \def\a{1.0}

  \pgfmathsetmacro{\bx}{\a}                
  \pgfmathsetmacro{\by}{0}
  \pgfmathsetmacro{\yx}{\a*cos(\angleY)}   
  \pgfmathsetmacro{\yy}{\a*sin(\angleY)}
  \pgfmathsetmacro{\zx}{0}                 
  \pgfmathsetmacro{\zy}{\a}

  \foreach \ix in {0,...,4}{
    \foreach \iy in {0,...,\Ny}{
      \foreach \iz in {0,...,\Nz}{
        \pgfmathsetmacro{\x}{\ix*\bx + \iy*\yx + \iz*\zx}
        \pgfmathsetmacro{\y}{\ix*\by + \iy*\yy + \iz*\zy}

        \ifnum\ix<\Nx
          \pgfmathsetmacro{\xnext}{(\ix+1)*\bx + \iy*\yx + \iz*\zx}
          \pgfmathsetmacro{\ynext}{(\ix+1)*\by + \iy*\yy + \iz*\zy}
          \draw[line width=0.6pt, gray] (\x,\y)--(\xnext,\ynext);
        \fi

        \ifnum\iy<\Ny
          \pgfmathsetmacro{\xnext}{\ix*\bx + (\iy+1)*\yx + \iz*\zx}
          \pgfmathsetmacro{\ynext}{\ix*\by + (\iy+1)*\yy + \iz*\zy}
          \draw[line width=0.6pt, gray] (\x,\y)--(\xnext,\ynext);
        \fi

        \ifnum\iz<\Nz
          \pgfmathsetmacro{\xnext}{\ix*\bx + \iy*\yx + (\iz+1)*\zx}
          \pgfmathsetmacro{\ynext}{\ix*\by + \iy*\yy + (\iz+1)*\zy}
          \draw[line width=0.6pt, gray] (\x,\y)--(\xnext,\ynext);
        \fi

        \filldraw[fill=black] (\x,\y) circle (0.02);
      }
    }
  }

\foreach \ix in {5}{
    \foreach \iy in {0,...,\Ny}{
      \foreach \iz in {0,...,\Nz}{
        \pgfmathsetmacro{\x}{\ix*\bx + \iy*\yx + \iz*\zx}
        \pgfmathsetmacro{\y}{\ix*\by + \iy*\yy + \iz*\zy}

        \ifnum\iy<\Ny
          \pgfmathsetmacro{\xnext}{\ix*\bx + (\iy+1)*\yx + \iz*\zx}
          \pgfmathsetmacro{\ynext}{\ix*\by + (\iy+1)*\yy + \iz*\zy}
          \draw[line width=0.6pt, dotted, gray] (\x,\y)--(\xnext,\ynext);
        \fi

        \ifnum\iz<\Nz
          \pgfmathsetmacro{\xnext}{\ix*\bx + \iy*\yx + (\iz+1)*\zx}
          \pgfmathsetmacro{\ynext}{\ix*\by + \iy*\yy + (\iz+1)*\zy}
          \draw[line width=0.6pt, dotted, gray] (\x,\y)--(\xnext,\ynext);
        \fi

        \filldraw[fill=black] (\x,\y) circle (0.02);
      }
    }
  }

  \foreach \ix in {2,...,4}{
    \foreach \iy in {0,...,\Ny}{
      \foreach \iz in {0,...,\Nz}{
        \pgfmathsetmacro{\x}{\ix*\bx + \iy*\yx + \iz*\zx}
        \pgfmathsetmacro{\y}{\ix*\by + \iy*\yy + \iz*\zy}

        \ifnum\iy<\Ny
          \pgfmathsetmacro{\xnext}{\ix*\bx + (\iy+1)*\yx + \iz*\zx}
          \pgfmathsetmacro{\ynext}{\ix*\by + (\iy+1)*\yy + \iz*\zy}
          \draw[line width=1.0pt, red] (\x,\y)--(\xnext,\ynext);
        \fi

        \ifnum\iz<\Nz
          \pgfmathsetmacro{\xnext}{\ix*\bx + \iy*\yx + (\iz+1)*\zx}
          \pgfmathsetmacro{\ynext}{\ix*\by + \iy*\yy + (\iz+1)*\zy}
          \draw[line width=1.0pt, red] (\x,\y)--(\xnext,\ynext);
        \fi
 
      }
    }
  }
  \vspace{0.2cm} 
\end{tikzpicture}
    \ee 
\end{subequations} 
We first perform the unitary transformation for both the Ising-dual and Majorana-dual gauge theories using the two layers of CNOT gates shown below. The arrow indicates the sequential application. Note that in the following schematics only a representative subset of the gates is depicted; the circuits should be understood as translationally invariant in planes perpendicular to the $x$-axis. For the Ising-dual gauge theory, this step is sufficient to decouple the Gauss-law operators if the space is infinite. 

\begin{subequations}
    
\be 

 \ee
\end{subequations} 
For the Majorana-dual gauge theory, we perform additional transformations using the following layers of CZ gates, sandwiched between a global basis change implemented by Hadamard gates.  
\begin{subequations} 
\be 
%
\ee
\end{subequations} 
To accommodate the periodic boundary condition, where the rightmost dotted edges are identified with the leftmost ones, additional operations are required. After the preceding transformations, the third Gauss-law operator in Fig.~\ref{fig:3d}(a) terminates on the boundary as follows. The part of the transformed Gauss-law operator acting on the boundary is precisely the Gauss-law operator in the 2D Majorana-dual gauge theory. 

\be 
%
 
\ee
Moreover, as in the 2D case, the three Gauss-law operators crossing the periodic boundary take a different form after the preceding transformations. For the Majorana-dual gauge theory, an additional unitary transformation, implemented by the circuit layers shown below, is required.

\begin{subequations}
\be 
%
\ee
\end{subequations}
Here the CNOT gates in the second layer act only within the rightmost cubes (indicated by the dotted boundary). 
After this transformation, the Gauss-law operators crossing the boundary in the Majorana-dual gauge theory become as follows. 
\begin{subequations}
\be 
%
\ee 
\end{subequations} 
The third Gauss-law operator sits on the rightmost plane and is again precisely the Gauss-law operator in the 2D Majorana-dual gauge theory. Interestingly, after the first step, the first two Gauss-law (plaquette) operators crossing the boundary in the Ising-dual gauge theory take the same form above while the third Gauss-law operator remains a plaquette operator on the rightmost plane, which is also the Gauss-law operator in the 2D Ising dual gauge theory. Therefore, we can apply the operations used in the 2D case to transform the Gauss-law operators on the rightmost plane so that they act only on the horizontal edges along the $y$-axis. Finally, we perform the subsequent SWAP operations that ensure all transformed Gauss-law operators act solely on edges perpendicular to the $x$-axis. 
\be 
\begin{tikzpicture}[scale=1.2, line cap=round, line join=round, baseline=(current bounding box.center)]

  \def\Nx{5}       
  \def\Ny{1}       
  \def\Nz{1}       
  \def\angleY{35}  

  \def\a{1.0}

  \pgfmathsetmacro{\bx}{\a}                
  \pgfmathsetmacro{\by}{0}
  \pgfmathsetmacro{\yx}{\a*cos(\angleY)}   
  \pgfmathsetmacro{\yy}{\a*sin(\angleY)}
  \pgfmathsetmacro{\zx}{0}                 
  \pgfmathsetmacro{\zy}{\a}

  \foreach \ix in {0,...,4}{
    \foreach \iy in {0,...,\Ny}{
      \foreach \iz in {0,...,\Nz}{
        \pgfmathsetmacro{\x}{\ix*\bx + \iy*\yx + \iz*\zx}
        \pgfmathsetmacro{\y}{\ix*\by + \iy*\yy + \iz*\zy}

        \ifnum\ix<\Nx
          \pgfmathsetmacro{\xnext}{(\ix+1)*\bx + \iy*\yx + \iz*\zx}
          \pgfmathsetmacro{\ynext}{(\ix+1)*\by + \iy*\yy + \iz*\zy}
          \draw[line width=0.6pt, gray] (\x,\y)--(\xnext,\ynext);
        \fi

        \ifnum\iy<\Ny
          \pgfmathsetmacro{\xnext}{\ix*\bx + (\iy+1)*\yx + \iz*\zx}
          \pgfmathsetmacro{\ynext}{\ix*\by + (\iy+1)*\yy + \iz*\zy}
          \draw[line width=0.6pt, gray] (\x,\y)--(\xnext,\ynext);
        \fi

        \ifnum\iz<\Nz
          \pgfmathsetmacro{\xnext}{\ix*\bx + \iy*\yx + (\iz+1)*\zx}
          \pgfmathsetmacro{\ynext}{\ix*\by + \iy*\yy + (\iz+1)*\zy}
          \draw[line width=0.6pt, gray] (\x,\y)--(\xnext,\ynext);
        \fi

        \filldraw[fill=black] (\x,\y) circle (0.02);
      }
    }
  }

  \foreach \ix in {5}{
    \foreach \iy in {0,...,\Ny}{
      \foreach \iz in {0,...,\Nz}{
        \pgfmathsetmacro{\x}{\ix*\bx + \iy*\yx + \iz*\zx}
        \pgfmathsetmacro{\y}{\ix*\by + \iy*\yy + \iz*\zy}

        \ifnum\iy<\Ny
          \pgfmathsetmacro{\xnext}{\ix*\bx + (\iy+1)*\yx + \iz*\zx}
          \pgfmathsetmacro{\ynext}{\ix*\by + (\iy+1)*\yy + \iz*\zy}
          \draw[line width=0.6pt, dotted, gray] (\x,\y)--(\xnext,\ynext);
        \fi

        \ifnum\iz<\Nz
          \pgfmathsetmacro{\xnext}{\ix*\bx + \iy*\yx + (\iz+1)*\zx}
          \pgfmathsetmacro{\ynext}{\ix*\by + \iy*\yy + (\iz+1)*\zy}
          \draw[line width=0.6pt, dotted, gray] (\x,\y)--(\xnext,\ynext);
        \fi

        \filldraw[fill=black] (\x,\y) circle (0.02);
      }
    }
  }

      \pgfmathsetmacro{\xA}{4.*\bx + 0*\yx+ 0.5*\zx}
      \pgfmathsetmacro{\yA}{4*\by + 0*\yy+ 0.5*\zy}
      \pgfmathsetmacro{\xB}{(4.5*\bx + 0.*\yx+ 1*\zx}
      \pgfmathsetmacro{\yB}{4.5*\by + 0*\yy+ 1*\zy}
      
       \draw[blue] (\xA, \yA) -- (\xB, \yB);

   \node[blue] at  (\xA, \yA) {$\times$};
       \node[blue] at  (\xB, \yB) {$\times$}; 

    \pgfmathsetmacro{\xA}{4.*\bx + 1*\yx+ 0.5*\zx}
      \pgfmathsetmacro{\yA}{4*\by + 1*\yy+ 0.5*\zy}
      \pgfmathsetmacro{\xB}{(4.5*\bx + 1.*\yx+ 1*\zx}
      \pgfmathsetmacro{\yB}{4.5*\by + 1*\yy+ 1*\zy}
      
       \draw[blue] (\xA, \yA) -- (\xB, \yB);
       \node[blue] at  (\xA, \yA) {$\times$};
       \node[blue] at  (\xB, \yB) {$\times$}; 

\end{tikzpicture}
\ee
After the above tedious transformations, the Gauss-law operators of both the Ising-dual and Majorana-dual gauge theories are transformed into identical operators acting on the edges on the planes perpendicular to the $x$-axis. As in the 2D case, the two unitary circuits can then be composed to construct a unitary mapping between the two gauge theories. 
 
So far, we have focused on the Gauss-law operators. To map the remaining operators in Fig.~\ref{fig:3d}(b) to those obtained via the Jordan–Wigner transformation of the folded Majorana chain, it is necessary to rearrange the images of these operators prior to the transformation shown in the last schematic. These omitted steps involve applying CNOT cascade operations along the $x$-axis, CNOT cascade operations in the rightmost plane perpendicular to the $x$-axis, and a translation along the $y$-axis within this plane. The reader can reconstruct the details if needed.

\end{appendix}





\bibliography{Bosonization.bib}

\end{document}